\documentstyle[12pt,axodraw,epsf]{article}
\begin{document}
\begin{flushright}
{DESY 01-049\\
April 2001}
\end{flushright}
\vspace{1.5cm}

\begin{center}
{\Large \bf Single Top Production \\ in $e^+ e^-$, $e^- e^-$, 
$\gamma e$ and $\gamma \gamma$ Collisions
\hfill\\}
\end{center}
\vspace{0.5cm}

\begin{center}
{E.~Boos$^{1,2}$, M.~Dubinin$^{1,2}$, A.~Pukhov$^1$,
M.~Sachwitz$^2$, H.J.~Schreiber$^2$ \\
\hfill\\
{\small \it $^1$Institute of Nuclear Physics, Moscow State University}\\
{\small \it 119899, Moscow, Russia} \\
 {\small \it $^2$DESY Zeuthen, D-15738 Zeuthen,
Germany} }
\end{center}

\vspace{1.0cm}
\begin{center}
{\bf Abstract}
\end{center}
\begin{quote}
Single top quark cross section evaluations for the complete sets of
tree-level diagrams in the 
$e^+ e^-$, $e^- e^-$, $\gamma e$ and $\gamma \gamma$ modes of the next 
linear collider
with unpolarized and polarized beams are performed within the Standard 
Model 
and beyond. 
From comparison of all possibilities we conclude that the process
$\gamma_+ e^-_L \to e^- t \bar b$ is extremely favoured due to large
cross sections, no $t \bar t$ background, high 
degrees of beam
polarization, and exceptional sensitivities to $V_{tb}$
and anomalous $Wtb$ couplings. Similar reasons favour the process
$e^- e^- \to e^- \nu_e \bar t b$ for probing
top quark properties despite a cosiderably lower cross section.
Less favourable are processes like 
$e^+ e^-, \gamma \gamma \to e^- \nu_e t \bar b$. 
Three processes were chosen to probe their sensitivity 
to anomalous $Wtb$ couplings, with best bounds found for  
$\gamma_+ e^-_L \to e^- t \bar b$ and $e^+_R e^-_R \to e^- \nu_e t \bar 
b$.
\end{quote}   

\newpage

\section{Introduction}

The most massive fermion of the Standard Model (SM), the top quark,
may provide unique possibilities to study the consistency of the SM and
the effects of new physics at energies of next linear colliders (LC). 
Precise 
calculations of the top production by colliding $e^\pm$ and $\gamma$ 
beams are very important for the
investigation of its couplings
to the gauge bosons and the Higgs bosons of the SM, of two Higgs
doublet models and their possible extensions.

In previous studies \cite{ee500} $e^+ e^- \to t \bar t$ pair production 
has been investigated in some details at c.m.s. energies near the 
threshold ($\sqrt{s} \sim$350 GeV), where the best
measurements of the top mass and width are expected, and also far 
above the threshold, 'in the continuum', where the top quark
currents and electric or magnetic moments 
can be probed with high precision. 
Single top production within the SM  has been
comparatively less studied. In the 2$\to$3 process the $e^+ e^- \to
W^+ t b$ single top cross section has been calculated below 
the $t \bar t$ threshold \cite{mele} while at higher energies 
\cite{jikia} most attention was focused on the
measurements of the Cabbibbo-Kobayashi-Maskawa (CKM) matrix 
element $V_{tb}$.
In these three particle final state approximations an important 
class of $t$-channel
diagrams with forward scattered electron has not been 
taken into account
\footnote{Early calculations of the $t$-channel subset diagrams
at TRISTAN energies can be found in \cite{shimizu0}, where the
denomination 'single top' has been introduced.}.
The calculation of the complete set of diagrams
for the four-fermion final state process 
$e^+ e^- \to e^- \bar \nu_e \bar t b$
has been performed for LEP2 energies in \cite{LEP2} and for
LC energies in \cite{sensitivity,kolodziej}.
It was concluded
that the event rate expected at LEP2 is so small that practically no
events would be observed. The analysis of \cite{sensitivity}
concerns the precision of $V_{tb}$ measurements at LC. 
There are also single top 
investigations for $\gamma e^-$ at LC \cite{ge_unpol} and $p \bar p$, $pp$ 
collisions at Tevatron and LHC \cite{Tevatron_LHC0} energies.

In this paper we evaluate cross sections 
for $e^+ e^-$, $e^- e^-$, $\gamma \gamma$ and $\gamma e^-$ collisions 
to the four fermion
final state $e^- \bar \nu_e t \bar b$, respectively, 
three fermion final state $\nu_e \bar t b$, taking into
account the complete sets of tree-level Feynman diagrams. 
Further, we consider besides unpolarized also polarized
initial state particle scattering, so full comparison 
of all initial state configurations of a linear collider can be performed.
Our analysis for single top production in polarized $\gamma e^-$ 
collisions goes beyond the study of \cite{ge_unpol} for the unpolarized 
case and demonstrates the advantages of beam polarization at LC. 
All calculations were performed by means
of the CompHEP package \cite{CompHEP}, after implementation of polarized
electron, positron and photon states. 
We consider the energy range from about 350 GeV
(where a non-negligible single top event rate at a high-luminosity
collider is expected) up to 1 TeV, planned for a first generation LC.

It has been emphasized in e.g. \cite{sensitivity} that single top
production is sensitive to the $V_{tb}$ matrix element since 
its rate is proportional to $|V_{tb}|^2$. Futhermore, if 
anomalous
effective operators for the $Wtb$ vertex are introduced,
the single top rate is sensitive to them, unlike the 
$t \bar t$ rate. This can be
easily understood in the {\it production $\times$ decay}
approximation for the $t \bar t$ amplitude and within the infinitely small   
top width approximation for the $s$-channel Breit-Wigner propagator
\begin{eqnarray}
\frac{\int |M(t\to Wb)|^2 d\Phi}{(q^2-m^2_{top})^2+m^2_{top}\,
\Gamma^2_{tot}}=Br(t \to Wb)\, \Gamma_{tot} \,
\frac{\pi}{m_{top}\, \Gamma_{tot}} \delta(q^2-m^2_{top}) \hspace{0.3cm}
\end{eqnarray}
where $Br(t \to Wb)= \Gamma(t \to Wb)/\Gamma_{tot}= \int |M(t\to Wb)|^2
\, d \Phi/\Gamma_{tot}$ is always close to 1,
and $\Gamma_{tot}$, sensitive to anomalies, cancels out.
By way of contrast,
simple counting of single top events could reveal signals
of anomalous $Wtb$ couplings,
and for a running
strategy of a linear collider it is worth to perform a comparative 
study of single top production for all possible collider options,
with unpolarized and polarized beams, and to point 
to the most appropriate collider mode.

Top quark studies in the six-fermion final states 
\cite{kuriharaballestrero} and their four particle final 
state approximation $W^+ b W^- \bar b$ \cite{ballestrero}
were focused on $t \bar t$ pair production. The case
of single top quark production was not considered there.

The paper is organised as follows. In section 2 we present
the single top quark cross sections for unpolarized and
polarized $e^+ e^-$, $e^- e^-$, $\gamma e^-$ and $\gamma \gamma$
collisions. In section 3 the anomalous
CP- and flavor conserving $Wtb$ operators of dimension 6 are introduced 
and,
as examples, 
possible bounds on anomalous couplings in $e^+_R e^-_R$, $e^-_L e^-_R$
and $\gamma_+ e^-_L$ collisions are discussed. Our conclusions are
presented in section 4.

\section{Cross sections in the $e^+ e^-$, $e^- e^-$, $\gamma e$ and
$\gamma \gamma$ modes of Linear Collider}

Single top quark production in $e^+ e^-$ and $\gamma \gamma$ 
collisions has to be extracted from
the complete sets of tree-level diagrams leading to the four-fermion 
final state $e^- \bar \nu_e \bar t b$, which is
the most general four-fermion case.
In unpolarized $e^+ e^-$ (Figs.1,2) and
$\gamma \gamma$ (Fig.3) collisions this 
final state appears either from $t \bar t$ pair production with 
the top decay into $e^- \bar \nu_e \bar b$ (see diagrams 3,4 in Fig.1
and diagrams 6,7 in Fig.3) or from single top production in
association with the $e^- \bar \nu_e \bar b$ system. After the elimination
of $t \bar t$ pair production by means of a subtraction 
procedure
(see section 2.1) all
diagrams in Figs.1-3 contribute to single top production. 
In $e^- e^-$ (Fig.4) and $\gamma e^-$ collisions (Fig.5) only
single top quark production is possible.

\unitlength 1cm
             
\begin{figure}[t]
\begin{center}
% diagrams for process e1,E1 -> N1,e1,t,B         
%\documentstyle[axodraw]{article}
%\begin{document}
{
\unitlength=0.7 pt
\SetScale{0.7}
\SetWidth{0.7}      % line    size control
\scriptsize    %  letter  size control
\noindent
{} \allowbreak
%  diagram # 1
\begin{picture}(95,99)(0,0)
\Text(15.0,90.0)[r]{$e$}
\ArrowLine(16.0,90.0)(37.0,80.0) 
\Text(15.0,70.0)[r]{$\bar{e}$}
\ArrowLine(37.0,80.0)(16.0,70.0) 
\Text(47.0,81.0)[b]{$\gamma,Z$}
\DashLine(37.0,80.0)(58.0,80.0){3.0} 
\Text(80.0,90.0)[l]{$e$}
\ArrowLine(58.0,80.0)(79.0,90.0) 
\Text(54.0,70.0)[r]{$e$}
\ArrowLine(58.0,60.0)(58.0,80.0) 
\Text(80.0,70.0)[l]{$\bar{\nu}_e$}
\ArrowLine(79.0,70.0)(58.0,60.0) 
\Text(54.0,50.0)[r]{$W^+$}
\DashArrowLine(58.0,60.0)(58.0,40.0){3.0} 
\Text(80.0,50.0)[l]{$\bar{b}$}
\ArrowLine(79.0,50.0)(58.0,40.0) 
\Text(80.0,30.0)[l]{$t$}
\ArrowLine(58.0,40.0)(79.0,30.0) 
\Text(47,0)[b] {diagr.1,2}
\end{picture} \ 
{} \qquad\allowbreak
%  diagram # 3
\begin{picture}(95,99)(0,0)
\Text(15.0,90.0)[r]{$e$}
\ArrowLine(16.0,90.0)(37.0,80.0) 
\Text(15.0,70.0)[r]{$\bar{e}$}
\ArrowLine(37.0,80.0)(16.0,70.0) 
\Text(47.0,81.0)[b]{$\gamma,Z$}
\DashLine(37.0,80.0)(58.0,80.0){3.0} 
\Text(80.0,90.0)[l]{$t$}
\ArrowLine(58.0,80.0)(79.0,90.0) 
\Text(54.0,70.0)[r]{$t$}
\ArrowLine(58.0,60.0)(58.0,80.0) 
\Text(80.0,70.0)[l]{$\bar{b}$}
\ArrowLine(79.0,70.0)(58.0,60.0) 
\Text(54.0,50.0)[r]{$W^+$}
\DashArrowLine(58.0,40.0)(58.0,60.0){3.0} 
\Text(80.0,50.0)[l]{$e$}
\ArrowLine(58.0,40.0)(79.0,50.0) 
\Text(80.0,30.0)[l]{$\bar{\nu}_e$}
\ArrowLine(79.0,30.0)(58.0,40.0) 
\Text(47,0)[b] {diagr.3,4}
\end{picture} \ 
{} \qquad\allowbreak
%  diagram # 5
\begin{picture}(95,99)(0,0)
\Text(15.0,90.0)[r]{$e$}
\ArrowLine(16.0,90.0)(37.0,80.0) 
\Text(15.0,70.0)[r]{$\bar{e}$}
\ArrowLine(37.0,80.0)(16.0,70.0) 
\Text(47.0,81.0)[b]{$\gamma,Z$}
\DashLine(37.0,80.0)(58.0,80.0){3.0} 
\Text(80.0,90.0)[l]{$\bar{b}$}
\ArrowLine(79.0,90.0)(58.0,80.0) 
\Text(54.0,70.0)[r]{$b$}
\ArrowLine(58.0,80.0)(58.0,60.0) 
\Text(80.0,70.0)[l]{$t$}
\ArrowLine(58.0,60.0)(79.0,70.0) 
\Text(54.0,50.0)[r]{$W^+$}
\DashArrowLine(58.0,40.0)(58.0,60.0){3.0} 
\Text(80.0,50.0)[l]{$e$}
\ArrowLine(58.0,40.0)(79.0,50.0) 
\Text(80.0,30.0)[l]{$\bar{\nu}_e$}
\ArrowLine(79.0,30.0)(58.0,40.0) 
\Text(47,0)[b] {diagr.5,6}
\end{picture} \ 
{} \qquad\allowbreak
%  diagram # 7
\begin{picture}(95,99)(0,0)
\Text(15.0,70.0)[r]{$e$}
\ArrowLine(16.0,70.0)(37.0,60.0) 
\Text(15.0,50.0)[r]{$\bar{e}$}
\ArrowLine(37.0,60.0)(16.0,50.0) 
\Text(37.0,62.0)[lb]{$\gamma,Z$}
\DashLine(37.0,60.0)(58.0,60.0){3.0} 
\Text(54.0,70.0)[r]{$$}
\DashArrowLine(58.0,80.0)(58.0,60.0){3.0} 
\Text(80.0,90.0)[l]{$e$}
\ArrowLine(58.0,80.0)(79.0,90.0) 
\Text(80.0,70.0)[l]{$\bar{\nu}_e$}
\ArrowLine(79.0,70.0)(58.0,80.0) 
\Text(54.0,50.0)[r]{$W^+$}
\DashArrowLine(58.0,60.0)(58.0,40.0){3.0} 
\Text(80.0,50.0)[l]{$\bar{b}$}
\ArrowLine(79.0,50.0)(58.0,40.0) 
\Text(80.0,30.0)[l]{$t$}
\ArrowLine(58.0,40.0)(79.0,30.0) 
\Text(47,0)[b] {diagr.7,8}
\end{picture} \ 
{} \qquad\allowbreak
%  diagram # 10
\begin{picture}(95,99)(0,0)
\Text(15.0,80.0)[r]{$e$}
\ArrowLine(16.0,80.0)(37.0,80.0) 
\Text(47.0,84.0)[b]{$W^+$}
\DashArrowLine(58.0,80.0)(37.0,80.0){3.0} 
\Text(80.0,90.0)[l]{$e$}
\ArrowLine(58.0,80.0)(79.0,90.0) 
\Text(80.0,70.0)[l]{$\bar{\nu}_e$}
\ArrowLine(79.0,70.0)(58.0,80.0) 
\Text(33.0,60.0)[r]{$\nu_e$}
\ArrowLine(37.0,80.0)(37.0,40.0) 
\Text(15.0,40.0)[r]{$\bar{e}$}
\ArrowLine(37.0,40.0)(16.0,40.0) 
\Text(47.0,44.0)[b]{$W^+$}
\DashArrowLine(37.0,40.0)(58.0,40.0){3.0} 
\Text(80.0,50.0)[l]{$\bar{b}$}
\ArrowLine(79.0,50.0)(58.0,40.0) 
\Text(80.0,30.0)[l]{$t$}
\ArrowLine(58.0,40.0)(79.0,30.0) 
\Text(47,0)[b] {diagr.9}
\end{picture} \ 
{} \qquad\allowbreak
%  diagram # 11
\begin{picture}(95,99)(0,0)
\Text(15.0,90.0)[r]{$e$}
\ArrowLine(16.0,90.0)(37.0,80.0) 
\Text(15.0,70.0)[r]{$\bar{e}$}
\ArrowLine(37.0,80.0)(16.0,70.0) 
\Text(47.0,81.0)[b]{$Z$}
\DashLine(37.0,80.0)(58.0,80.0){3.0} 
\Text(80.0,90.0)[l]{$\bar{\nu}_e$}
\ArrowLine(79.0,90.0)(58.0,80.0) 
\Text(54.0,70.0)[r]{$\nu_e$}
\ArrowLine(58.0,80.0)(58.0,60.0) 
\Text(80.0,70.0)[l]{$e$}
\ArrowLine(58.0,60.0)(79.0,70.0) 
\Text(54.0,50.0)[r]{$W^+$}
\DashArrowLine(58.0,60.0)(58.0,40.0){3.0} 
\Text(80.0,50.0)[l]{$\bar{b}$}
\ArrowLine(79.0,50.0)(58.0,40.0) 
\Text(80.0,30.0)[l]{$t$}
\ArrowLine(58.0,40.0)(79.0,30.0) 
\Text(47,0)[b] {diagr.10}
\end{picture} \ 
}
%\end{document}
\end{center}
\caption{$s$-channel  $CC10$ diagrams for the process $e^- e^+ \to e^- 
\bar \nu_e t \bar b$, unpolarized beams}
\end{figure}
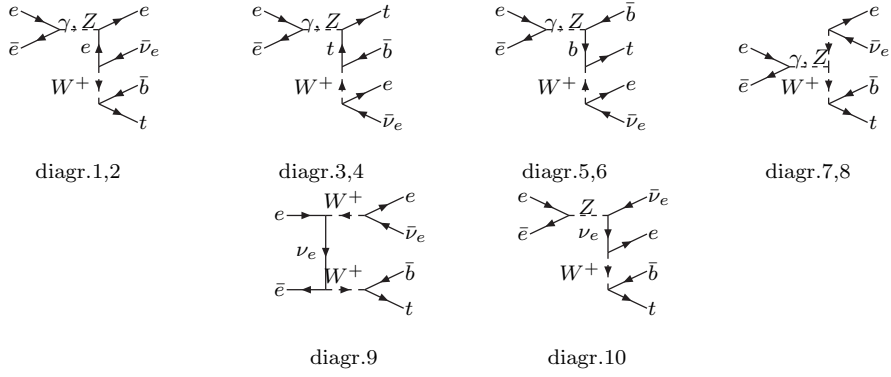
% Fig 1

\begin{figure}[h]
\begin{center}
% diagrams for process e1,E1 -> N1,e1,t,B         
%\documentstyle[axodraw]{article}
%\begin{document}
{
\unitlength=0.7 pt
\SetScale{0.7}
\SetWidth{0.7}      % line    size control
\scriptsize    %  letter  size control
\noindent
{} \allowbreak
%  diagram # 2
\begin{picture}(95,99)(0,0)
\Text(15.0,80.0)[r]{$e$}
\ArrowLine(16.0,80.0)(37.0,80.0) 
\Line(37.0,80.0)(58.0,80.0) 
\Text(80.0,90.0)[l]{$e$}
\ArrowLine(58.0,80.0)(79.0,90.0) 
\Text(36.0,70.0)[r]{$\gamma,Z$}
\DashLine(37.0,80.0)(37.0,60.0){3.0} 
\Text(15.0,60.0)[r]{$\bar{e}$}
\ArrowLine(37.0,60.0)(16.0,60.0) 
\Text(47.0,64.0)[b]{$e$}
\ArrowLine(58.0,60.0)(37.0,60.0) 
\Text(80.0,70.0)[l]{$\bar{\nu}_e$}
\ArrowLine(79.0,70.0)(58.0,60.0) 
\Text(54.0,50.0)[r]{$W^+$}
\DashArrowLine(58.0,60.0)(58.0,40.0){3.0} 
\Text(80.0,50.0)[l]{$\bar{b}$}
\ArrowLine(79.0,50.0)(58.0,40.0) 
\Text(80.0,30.0)[l]{$t$}
\ArrowLine(58.0,40.0)(79.0,30.0) 
\Text(47,0)[b] {diagr.1,2}
\end{picture} \ 
{} \qquad\allowbreak
%  diagram # 4
\begin{picture}(95,99)(0,0)
\Text(15.0,90.0)[r]{$e$}
\ArrowLine(16.0,90.0)(58.0,90.0) 
\Text(80.0,90.0)[l]{$e$}
\ArrowLine(58.0,90.0)(79.0,90.0) 
\Text(57.0,80.0)[r]{$\gamma,Z$}
\DashLine(58.0,90.0)(58.0,70.0){3.0} 
\Text(80.0,70.0)[l]{$t$}
\ArrowLine(58.0,70.0)(79.0,70.0) 
\Text(54.0,60.0)[r]{$t$}
\ArrowLine(58.0,50.0)(58.0,70.0) 
\Text(80.0,50.0)[l]{$\bar{b}$}
\ArrowLine(79.0,50.0)(58.0,50.0) 
\Text(54.0,40.0)[r]{$W^+$}
\DashArrowLine(58.0,30.0)(58.0,50.0){3.0} 
\Text(15.0,30.0)[r]{$\bar{e}$}
\ArrowLine(58.0,30.0)(16.0,30.0) 
\Text(80.0,30.0)[l]{$\bar{\nu}_e$}
\ArrowLine(79.0,30.0)(58.0,30.0) 
\Text(47,0)[b] {diagr.3,4}
\end{picture} \ 
{} \qquad\allowbreak
%  diagram # 6
\begin{picture}(95,99)(0,0)
\Text(15.0,90.0)[r]{$e$}
\ArrowLine(16.0,90.0)(58.0,90.0) 
\Text(80.0,90.0)[l]{$e$}
\ArrowLine(58.0,90.0)(79.0,90.0) 
\Text(57.0,80.0)[r]{$\gamma,Z$}
\DashLine(58.0,90.0)(58.0,70.0){3.0} 
\Text(80.0,70.0)[l]{$\bar{b}$}
\ArrowLine(79.0,70.0)(58.0,70.0) 
\Text(54.0,60.0)[r]{$b$}
\ArrowLine(58.0,70.0)(58.0,50.0) 
\Text(80.0,50.0)[l]{$t$}
\ArrowLine(58.0,50.0)(79.0,50.0) 
\Text(54.0,40.0)[r]{$W^+$}
\DashArrowLine(58.0,30.0)(58.0,50.0){3.0} 
\Text(15.0,30.0)[r]{$\bar{e}$}
\ArrowLine(58.0,30.0)(16.0,30.0) 
\Text(80.0,30.0)[l]{$\bar{\nu}_e$}
\ArrowLine(79.0,30.0)(58.0,30.0) 
\Text(47,0)[b] {diagr.5,6}
\end{picture} \ 
{} \qquad\allowbreak
%  diagram # 8
\begin{picture}(95,99)(0,0)
\Text(15.0,80.0)[r]{$e$}
\ArrowLine(16.0,80.0)(37.0,80.0) 
\Line(37.0,80.0)(58.0,80.0) 
\Text(80.0,90.0)[l]{$e$}
\ArrowLine(58.0,80.0)(79.0,90.0) 
\Text(36.0,70.0)[r]{$\gamma,Z$}
\DashLine(37.0,80.0)(37.0,60.0){3.0} 
\Text(47.0,64.0)[b]{$W^+$}
\DashArrowLine(37.0,60.0)(58.0,60.0){3.0} 
\Text(80.0,70.0)[l]{$\bar{b}$}
\ArrowLine(79.0,70.0)(58.0,60.0) 
\Text(80.0,50.0)[l]{$t$}
\ArrowLine(58.0,60.0)(79.0,50.0) 
\Text(33.0,50.0)[r]{$W^+$}
\DashArrowLine(37.0,40.0)(37.0,60.0){3.0} 
\Text(15.0,40.0)[r]{$\bar{e}$}
\ArrowLine(37.0,40.0)(16.0,40.0) 
\Line(37.0,40.0)(58.0,40.0) 
\Text(80.0,30.0)[l]{$\bar{\nu}_e$}
\ArrowLine(79.0,30.0)(58.0,40.0) 
\Text(47,0)[b] {diagr.7,8}
\end{picture} \ 
{} \qquad\allowbreak
%  diagram # 9
\begin{picture}(95,99)(0,0)
\Text(15.0,80.0)[r]{$e$}
\ArrowLine(16.0,80.0)(37.0,80.0) 
\Text(47.0,84.0)[b]{$\nu_e$}
\ArrowLine(37.0,80.0)(58.0,80.0) 
\Text(80.0,90.0)[l]{$e$}
\ArrowLine(58.0,80.0)(79.0,90.0) 
\Text(54.0,70.0)[r]{$W^+$}
\DashArrowLine(58.0,80.0)(58.0,60.0){3.0} 
\Text(80.0,70.0)[l]{$\bar{b}$}
\ArrowLine(79.0,70.0)(58.0,60.0) 
\Text(80.0,50.0)[l]{$t$}
\ArrowLine(58.0,60.0)(79.0,50.0) 
\Text(33.0,60.0)[r]{$W^+$}
\DashArrowLine(37.0,40.0)(37.0,80.0){3.0} 
\Text(15.0,40.0)[r]{$\bar{e}$}
\ArrowLine(37.0,40.0)(16.0,40.0) 
\Line(37.0,40.0)(58.0,40.0) 
\Text(80.0,30.0)[l]{$\bar{\nu}_e$}
\ArrowLine(79.0,30.0)(58.0,40.0) 
\Text(47,0)[b] {diagr.9}
\end{picture} \ 
{} \qquad\allowbreak
%  diagram # 12
\begin{picture}(95,99)(0,0)
\Text(15.0,80.0)[r]{$e$}
\ArrowLine(16.0,80.0)(37.0,80.0) 
\Line(37.0,80.0)(58.0,80.0) 
\Text(80.0,90.0)[l]{$e$}
\ArrowLine(58.0,80.0)(79.0,90.0) 
\Text(36.0,70.0)[r]{$Z$}
\DashLine(37.0,80.0)(37.0,60.0){3.0} 
\Line(37.0,60.0)(58.0,60.0) 
\Text(80.0,70.0)[l]{$\bar{\nu}_e$}
\ArrowLine(79.0,70.0)(58.0,60.0) 
\Text(33.0,50.0)[r]{$\nu_e$}
\ArrowLine(37.0,60.0)(37.0,40.0) 
\Text(15.0,40.0)[r]{$\bar{e}$}
\ArrowLine(37.0,40.0)(16.0,40.0) 
\Text(47.0,44.0)[b]{$W^+$}
\DashArrowLine(37.0,40.0)(58.0,40.0){3.0} 
\Text(80.0,50.0)[l]{$\bar{b}$}
\ArrowLine(79.0,50.0)(58.0,40.0) 
\Text(80.0,30.0)[l]{$t$}
\ArrowLine(58.0,40.0)(79.0,30.0) 
\Text(47,0)[b] {diagr.10}
\end{picture} \ 
}
%\end{document}
\end{center}
\caption{$t$-channel $CC10$ diagrams for the process $e^- e^+ \to e^- \bar 
\nu_e t \bar b$, unpolarized beams}
\end{figure}
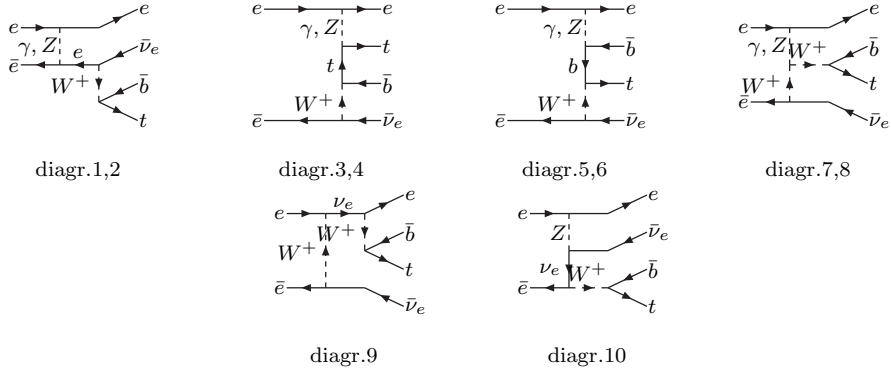
% Fig 2    

Appropriate beam polarization gives the possibility to exclude 
$t \bar t$ pair production in the $e^+ e^-$ case and 
allows for only single top quark production. Such a possibility
does not exist for $\gamma \gamma$ collisions, whatever polarizations
one assumes.

Table 1 summarizes the various possibilities expected for a linear 
collider.
Discussions of single top quark production cross sections for all 
collision modes offered by a LC are presented in the following.

\begin{figure}[h!]
\begin{center}
\input{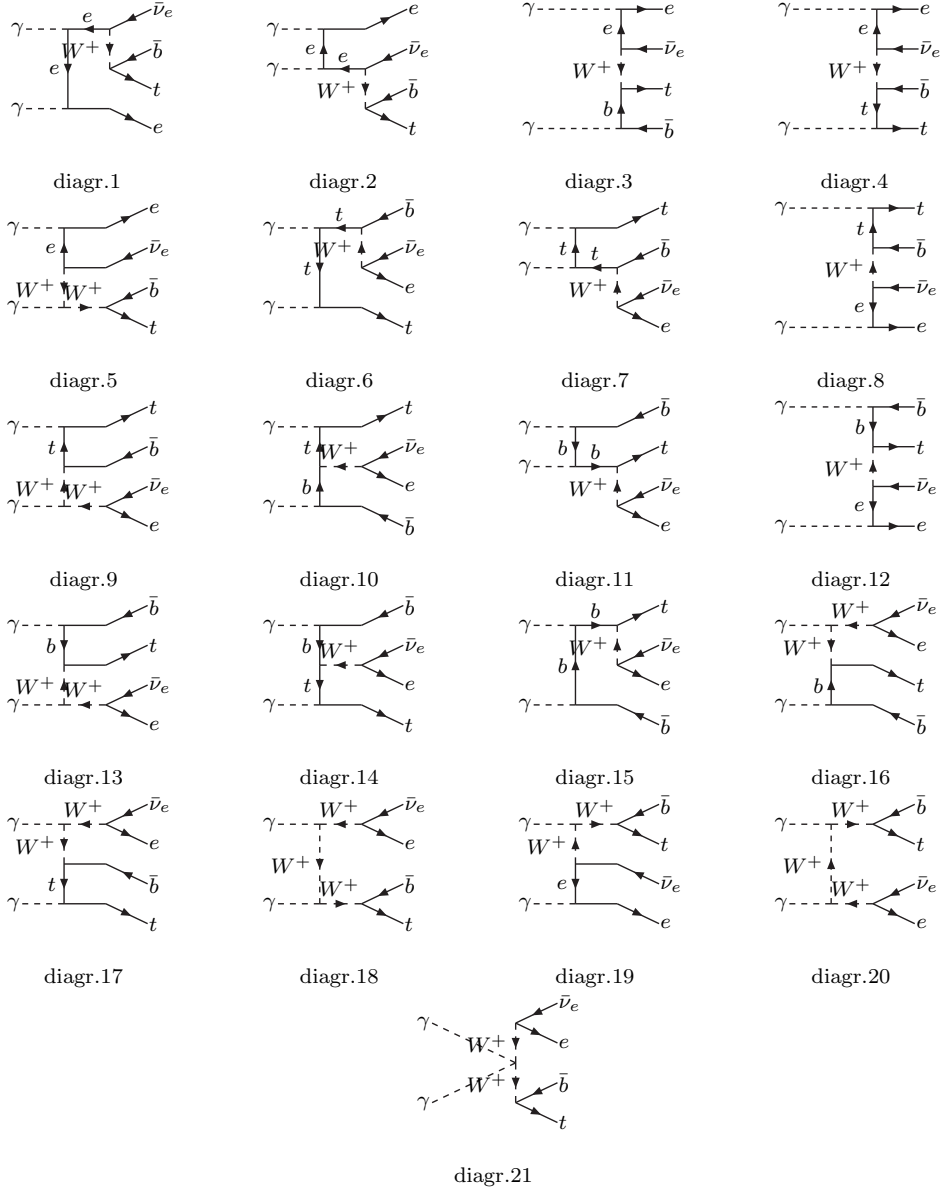}
\end{center}  
\caption{Diagrams for the process $\gamma \gamma \to e^- \nu_e \bar t
b$}
\end{figure}
% Fig 3

\begin{table}[t]
\begin{center}
\begin{tabular}{|c|c|c|c|c|c|}
\hline
 beams & No. of& polarization & $t \bar t$ & $\sigma_{single \;
top}$, fb &
$\sigma_{single \; top}$, fb \\
       & diagrams & and subset & production &  $\sqrt{s}=$0.5 TeV &
                                     $\sqrt{s}=$1 TeV \\
\hline
 $e^- e^+$ & 20 & unpol       &  yes &  3.1 & 6.7\\
           & 10 & unpol,$s$-ch.& yes & 2.3 & 2.3 \\
           & 10 & unpol,$t$-ch.& no                &  0.8 & 4.4 \\
           & 20 & LR          &  yes & 10.0& 16.9\\
           & 11 & RL          &  yes & 1.7 & 1.0 \\
           &  9 & RR          &  no             &  1.0 & 8.1\\
           &  2 & LL          &  no                & - & - \\
\hline
 $e^- e^-$ & 20 & unpol       &  no              & 1.7 & 9.1 \\
           & 20 & LL          &  no              & 2.6 & 19.1 \\
           & 11 & LR          &  no                & 2.1 & 14.0 \\
           & 11 & RL          &  no                & 2.1 & 14.0 \\
           &  4 & RR          &  no                & 0.02 & 0.96 \\
\hline
$\gamma e^-$ &4 & unpol       &  no                & 30.3 & 67.6 \\
             &4 & $-$L    & no                &  38.9 & 121.3 \\
             &4 & +L    & no                &  94.3 & 174.7 \\
\hline
$\gamma \gamma$& 21& unpol    & yes  & 9.2 & 18.8\\
           & 21 &(++)         & yes  & 11.1 & 19.2\\
           & 21 &($-$$-$)     & yes  & 7.9 & 15.7\\
           & 21 &(+$-$) or ($-$+) & yes & 8.5& 19.2\\
\hline
       
\end{tabular}
\end{center}
\caption{Top quark production by $e^+$, $e^-$ and $\gamma$ beams with
various polarizations. The single top quark production cross sections 
(with $m_{top}=$175 GeV) are
given for the channels
 $e^- e^+, \gamma \gamma \to e^- \bar \nu_e t \bar b$,
 $e^- e^- \to e^- \nu_e \bar t b$ and $\gamma e^- \to \nu_e \bar t b$
at $\sqrt{s}=$0.5 and 1.0 TeV.}
\end{table}

\subsection{Single top production in $e^+ e^-$ collisions with 
unpolarized and polarized beams}

In order to calculate at complete tree level the 
single top cross section in the 
reaction $e^- e^+ \to e^- \bar \nu_e t \bar b$ 
\footnote{The final states $\mu \nu_{\mu} t \bar b$ 
and $\tau \nu_{\tau} t \bar b$ produce single top events with the 
same rate as the  $CC10$ $s$-channel diagrams of Fig.1.
The definitions of $CC10$ and $CC20$ are given in \cite{Bardin}.}  
with unpolarized beams,
all 20 Feynman diagrams of Figs.1 and 2 have to be taken 
into account to ensure gauge invariance. The diagrams in these 
figures form two minimal \cite{boos-ohl} gauge invariant 
subsets. 
The energy dependence of the cross sections for the entire 
$e^- e^+ \to e^- \bar \nu_e t \bar b$
reaction and the $s$- and $t$-channel subsets is shown in Fig.6a.  
Most of the $e^- e^+ \to e^- \bar \nu_e t \bar b$ cross section 
comes from the $s$-channel diagrams which in turn are dominated by
$t \bar t$ pair production (diagrams 3,4 in Fig.1).
This contribution has to be 
subtracted from the total event rate in a gauge invariant manner
to get the single top rate. 

We define the single top cross section as the difference 
of the complete tree-level (CTL) contribution and the Breit-Wigner (BW) 
resonance contribution
$\int dM_{e \nu b} \, (d\sigma^{CTL}/dM_{e \nu b}-d\sigma^{BW}/dM_{e \nu 
b})$.  

However, for simplicity we applied a cut on the $e \nu b$ invariant mass
around the top quark pole \cite{sensitivity, LHC00} as an equivalent of 
the BW subtraction procedure
\begin{equation}
\sigma=\int\limits_{M_{min}}^{m_{top}-\Delta} dM_{e \nu b} 
\frac{d\sigma^{CTL}}{dM_{e \nu b}}
      +\int\limits^{M_{max}}_{m_{top}+\Delta} dM_{e \nu b} 
\frac{d\sigma^{CTL}}{dM_{e \nu b}},
\end{equation}
where the value of $\Delta$ is adjusted to compensate the small amount of 
discarded single top events inside the interval $m_{top}-\Delta \leq 
m_{top} \leq m_{top}+\Delta$ by the remaining top Breit-Wigner tails 
outside. The cross sections obtained using (2) have been compared
with those based on a subtraction of $t \bar t$ cross sections
from a fit of a Breit-Wigner function superimposed with a polynomial 
to $M_{e \nu_e b}$. Both numbers 
agree very well 
if $\Delta$ is taken to be 20 GeV. This value of $\Delta$ is much larger 
than  
an intuitively expected one of the 
order of the top quark width, which would lead to large contributions of 
surviving $t \bar t$ events. Obviously, the procedure applied is gauge 
invariant. The resulting single top cross section 
for unpolarized $e^+ e^-$ collisions
is shown in Fig.7 (solid curve). Below the $t \bar t$ threshold
it is less than 1 fb and increases up to 7 fb at
$\sqrt{s}=$1 TeV.
\begin{figure}[h!]
\begin{center}
% diagrams for process e1,e1 -> e1,n1,T,b         
%\documentstyle[axodraw]{article}
%\begin{document}
{
\unitlength=0.7 pt
\SetScale{0.7}
\SetWidth{0.7}      % line    size control
\scriptsize    %  letter  size control
{} \qquad\allowbreak
%  diagram # 1
\begin{picture}(95,99)(0,0)
\Text(15.0,80.0)[r]{$e$}
\ArrowLine(16.0,80.0)(37.0,80.0) 
\Text(47.0,84.0)[b]{$e$}
\ArrowLine(37.0,80.0)(58.0,80.0) 
\Text(80.0,90.0)[l]{$\nu_e$}
\ArrowLine(58.0,80.0)(79.0,90.0) 
\Text(54.0,70.0)[r]{$W^+$}
\DashArrowLine(58.0,60.0)(58.0,80.0){3.0} 
\Text(80.0,70.0)[l]{$b$}
\ArrowLine(58.0,60.0)(79.0,70.0) 
\Text(80.0,50.0)[l]{$\bar{t}$}
\ArrowLine(79.0,50.0)(58.0,60.0) 
\Text(36.0,60.0)[r]{$\gamma,Z$}
\DashLine(37.0,80.0)(37.0,40.0){3.0} 
\Text(15.0,40.0)[r]{$e$}
\ArrowLine(16.0,40.0)(37.0,40.0) 
\Line(37.0,40.0)(58.0,40.0) 
\Text(80.0,30.0)[l]{$e$}
\ArrowLine(58.0,40.0)(79.0,30.0) 
\Text(47,0)[b] {diagr.1,2}
\end{picture} \ 
{} \qquad\allowbreak
%  diagram # 2
\begin{picture}(95,99)(0,0)
\Text(15.0,80.0)[r]{$e$}
\ArrowLine(16.0,80.0)(37.0,80.0) 
\Line(37.0,80.0)(58.0,80.0) 
\Text(80.0,90.0)[l]{$e$}
\ArrowLine(58.0,80.0)(79.0,90.0) 
\Text(36.0,70.0)[r]{$\gamma,Z$}
\DashLine(37.0,80.0)(37.0,60.0){3.0} 
\Text(15.0,60.0)[r]{$e$}
\ArrowLine(16.0,60.0)(37.0,60.0) 
\Text(47.0,64.0)[b]{$e$}
\ArrowLine(37.0,60.0)(58.0,60.0) 
\Text(80.0,70.0)[l]{$\nu_e$}
\ArrowLine(58.0,60.0)(79.0,70.0) 
\Text(54.0,50.0)[r]{$W^+$}
\DashArrowLine(58.0,40.0)(58.0,60.0){3.0} 
\Text(80.0,50.0)[l]{$b$}
\ArrowLine(58.0,40.0)(79.0,50.0) 
\Text(80.0,30.0)[l]{$\bar{t}$}
\ArrowLine(79.0,30.0)(58.0,40.0) 
\Text(47,0)[b] {diagr.3,4}
\end{picture} \ 
{} \qquad\allowbreak
%  diagram # 3
\begin{picture}(95,99)(0,0)
\Text(15.0,90.0)[r]{$e$}
\ArrowLine(16.0,90.0)(58.0,90.0) 
\Text(80.0,90.0)[l]{$e$}
\ArrowLine(58.0,90.0)(79.0,90.0) 
\Text(57.0,80.0)[r]{$\gamma,Z$}
\DashLine(58.0,90.0)(58.0,70.0){3.0} 
\Text(80.0,70.0)[l]{$\bar{t}$}
\ArrowLine(79.0,70.0)(58.0,70.0) 
\Text(54.0,60.0)[r]{$t$}
\ArrowLine(58.0,70.0)(58.0,50.0) 
\Text(80.0,50.0)[l]{$b$}
\ArrowLine(58.0,50.0)(79.0,50.0) 
\Text(54.0,40.0)[r]{$W^+$}
\DashArrowLine(58.0,50.0)(58.0,30.0){3.0} 
\Text(15.0,30.0)[r]{$e$}
\ArrowLine(16.0,30.0)(58.0,30.0) 
\Text(80.0,30.0)[l]{$\nu_e$}
\ArrowLine(58.0,30.0)(79.0,30.0) 
\Text(47,0)[b] {diagr.5,6}
\end{picture} \ 
{} \qquad\allowbreak
%  diagram # 4
\begin{picture}(95,99)(0,0)
\Text(15.0,90.0)[r]{$e$}
\ArrowLine(16.0,90.0)(58.0,90.0) 
\Text(80.0,90.0)[l]{$e$}
\ArrowLine(58.0,90.0)(79.0,90.0) 
\Text(57.0,80.0)[r]{$\gamma,Z$}
\DashLine(58.0,90.0)(58.0,70.0){3.0} 
\Text(80.0,70.0)[l]{$b$}
\ArrowLine(58.0,70.0)(79.0,70.0) 
\Text(54.0,60.0)[r]{$b$}
\ArrowLine(58.0,50.0)(58.0,70.0) 
\Text(80.0,50.0)[l]{$\bar{t}$}
\ArrowLine(79.0,50.0)(58.0,50.0) 
\Text(54.0,40.0)[r]{$W^+$}
\DashArrowLine(58.0,50.0)(58.0,30.0){3.0} 
\Text(15.0,30.0)[r]{$e$}
\ArrowLine(16.0,30.0)(58.0,30.0) 
\Text(80.0,30.0)[l]{$\nu_e$}
\ArrowLine(58.0,30.0)(79.0,30.0) 
\Text(47,0)[b] {diagr.7,8}
\end{picture} \ 
{} \qquad\allowbreak
%  diagram # 5
\begin{picture}(95,99)(0,0)
\Text(15.0,80.0)[r]{$e$}
\ArrowLine(16.0,80.0)(37.0,80.0) 
\Line(37.0,80.0)(58.0,80.0) 
\Text(80.0,90.0)[l]{$e$}
\ArrowLine(58.0,80.0)(79.0,90.0) 
\Text(36.0,70.0)[r]{$\gamma,Z$}
\DashLine(37.0,80.0)(37.0,60.0){3.0} 
\Text(47.0,64.0)[b]{$W^+$}
\DashArrowLine(58.0,60.0)(37.0,60.0){3.0} 
\Text(80.0,70.0)[l]{$b$}
\ArrowLine(58.0,60.0)(79.0,70.0) 
\Text(80.0,50.0)[l]{$\bar{t}$}
\ArrowLine(79.0,50.0)(58.0,60.0) 
\Text(33.0,50.0)[r]{$W^+$}
\DashArrowLine(37.0,60.0)(37.0,40.0){3.0} 
\Text(15.0,40.0)[r]{$e$}
\ArrowLine(16.0,40.0)(37.0,40.0) 
\Line(37.0,40.0)(58.0,40.0) 
\Text(80.0,30.0)[l]{$\nu_e$}
\ArrowLine(58.0,40.0)(79.0,30.0) 
\Text(47,0)[b] {diagr.9,10}
\end{picture} \ 
{} \qquad\allowbreak
%  diagram # 6
\begin{picture}(95,99)(0,0)
\Text(15.0,80.0)[r]{$e$}
\ArrowLine(16.0,80.0)(37.0,80.0) 
\Line(37.0,80.0)(58.0,80.0) 
\Text(80.0,90.0)[l]{$\nu_e$}
\ArrowLine(58.0,80.0)(79.0,90.0) 
\Text(33.0,70.0)[r]{$W^+$}
\DashArrowLine(37.0,60.0)(37.0,80.0){3.0} 
\Line(37.0,60.0)(58.0,60.0) 
\Text(80.0,70.0)[l]{$e$}
\ArrowLine(58.0,60.0)(79.0,70.0) 
\Text(33.0,50.0)[r]{$\nu_e$}
\ArrowLine(37.0,40.0)(37.0,60.0) 
\Text(15.0,40.0)[r]{$e$}
\ArrowLine(16.0,40.0)(37.0,40.0) 
\Text(47.0,44.0)[b]{$W^+$}
\DashArrowLine(58.0,40.0)(37.0,40.0){3.0} 
\Text(80.0,50.0)[l]{$b$}
\ArrowLine(58.0,40.0)(79.0,50.0) 
\Text(80.0,30.0)[l]{$\bar{t}$}
\ArrowLine(79.0,30.0)(58.0,40.0) 
\Text(47,0)[b] {diagr.11}
\end{picture} \ 
{} \qquad\allowbreak
%  diagram # 7
\begin{picture}(95,99)(0,0)
\Text(15.0,90.0)[r]{$e$}
\ArrowLine(16.0,90.0)(58.0,90.0) 
\Text(80.0,90.0)[l]{$\nu_e$}
\ArrowLine(58.0,90.0)(79.0,90.0) 
\Text(54.0,80.0)[r]{$W^+$}
\DashArrowLine(58.0,70.0)(58.0,90.0){3.0} 
\Text(80.0,70.0)[l]{$\bar{t}$}
\ArrowLine(79.0,70.0)(58.0,70.0) 
\Text(54.0,60.0)[r]{$b$}
\ArrowLine(58.0,70.0)(58.0,50.0) 
\Text(80.0,50.0)[l]{$b$}
\ArrowLine(58.0,50.0)(79.0,50.0) 
\Text(57.0,40.0)[r]{$\gamma,Z$}
\DashLine(58.0,50.0)(58.0,30.0){3.0} 
\Text(15.0,30.0)[r]{$e$}
\ArrowLine(16.0,30.0)(58.0,30.0) 
\Text(80.0,30.0)[l]{$e$}
\ArrowLine(58.0,30.0)(79.0,30.0) 
\Text(47,0)[b] {diagr.12,13}
\end{picture} \ 
{} \qquad\allowbreak
%  diagram # 9
\begin{picture}(95,99)(0,0)
\Text(15.0,90.0)[r]{$e$}
\ArrowLine(16.0,90.0)(58.0,90.0) 
\Text(80.0,90.0)[l]{$\nu_e$}
\ArrowLine(58.0,90.0)(79.0,90.0) 
\Text(54.0,80.0)[r]{$W^+$}
\DashArrowLine(58.0,70.0)(58.0,90.0){3.0} 
\Text(80.0,70.0)[l]{$b$}
\ArrowLine(58.0,70.0)(79.0,70.0) 
\Text(54.0,60.0)[r]{$t$}
\ArrowLine(58.0,50.0)(58.0,70.0) 
\Text(80.0,50.0)[l]{$\bar{t}$}
\ArrowLine(79.0,50.0)(58.0,50.0) 
\Text(57.0,40.0)[r]{$\gamma,Z$}
\DashLine(58.0,50.0)(58.0,30.0){3.0} 
\Text(15.0,30.0)[r]{$e$}
\ArrowLine(16.0,30.0)(58.0,30.0) 
\Text(80.0,30.0)[l]{$e$}
\ArrowLine(58.0,30.0)(79.0,30.0) 
\Text(47,0)[b] {diagr.14,15}
\end{picture} \ 
{} \qquad\allowbreak
%  diagram # 11
\begin{picture}(95,99)(0,0)
\Text(15.0,80.0)[r]{$e$}
\ArrowLine(16.0,80.0)(37.0,80.0) 
\Line(37.0,80.0)(58.0,80.0) 
\Text(80.0,90.0)[l]{$\nu_e$}
\ArrowLine(58.0,80.0)(79.0,90.0) 
\Text(33.0,70.0)[r]{$W^+$}
\DashArrowLine(37.0,60.0)(37.0,80.0){3.0} 
\Text(47.0,64.0)[b]{$W^+$}
\DashArrowLine(58.0,60.0)(37.0,60.0){3.0} 
\Text(80.0,70.0)[l]{$b$}
\ArrowLine(58.0,60.0)(79.0,70.0) 
\Text(80.0,50.0)[l]{$\bar{t}$}
\ArrowLine(79.0,50.0)(58.0,60.0) 
\Text(36.0,50.0)[r]{$\gamma,Z$}
\DashLine(37.0,60.0)(37.0,40.0){3.0} 
\Text(15.0,40.0)[r]{$e$}
\ArrowLine(16.0,40.0)(37.0,40.0) 
\Line(37.0,40.0)(58.0,40.0) 
\Text(80.0,30.0)[l]{$e$}
\ArrowLine(58.0,40.0)(79.0,30.0) 
\Text(47,0)[b] {diagr.16,17}
\end{picture} \ 
{} \qquad\allowbreak
%  diagram # 13
\begin{picture}(95,99)(0,0)
\Text(15.0,80.0)[r]{$e$}
\ArrowLine(16.0,80.0)(37.0,80.0) 
\Text(47.0,84.0)[b]{$W^+$}
\DashArrowLine(58.0,80.0)(37.0,80.0){3.0} 
\Text(80.0,90.0)[l]{$b$}
\ArrowLine(58.0,80.0)(79.0,90.0) 
\Text(80.0,70.0)[l]{$\bar{t}$}
\ArrowLine(79.0,70.0)(58.0,80.0) 
\Text(33.0,70.0)[r]{$\nu_e$}
\ArrowLine(37.0,80.0)(37.0,60.0) 
\Line(37.0,60.0)(58.0,60.0) 
\Text(80.0,50.0)[l]{$e$}
\ArrowLine(58.0,60.0)(79.0,50.0) 
\Text(33.0,50.0)[r]{$W^+$}
\DashArrowLine(37.0,60.0)(37.0,40.0){3.0} 
\Text(15.0,40.0)[r]{$e$}
\ArrowLine(16.0,40.0)(37.0,40.0) 
\Line(37.0,40.0)(58.0,40.0) 
\Text(80.0,30.0)[l]{$\nu_e$}
\ArrowLine(58.0,40.0)(79.0,30.0) 
\Text(47,0)[b] {diagr.18}
\end{picture} \ 
{} \qquad\allowbreak
%  diagram # 14
\begin{picture}(95,99)(0,0)
\Text(15.0,80.0)[r]{$e$}
\ArrowLine(16.0,80.0)(37.0,80.0) 
\Text(47.0,84.0)[b]{$W^+$}
\DashArrowLine(58.0,80.0)(37.0,80.0){3.0} 
\Text(80.0,90.0)[l]{$b$}
\ArrowLine(58.0,80.0)(79.0,90.0) 
\Text(80.0,70.0)[l]{$\bar{t}$}
\ArrowLine(79.0,70.0)(58.0,80.0) 
\Text(33.0,70.0)[r]{$\nu_e$}
\ArrowLine(37.0,80.0)(37.0,60.0) 
\Line(37.0,60.0)(58.0,60.0) 
\Text(80.0,50.0)[l]{$\nu_e$}
\ArrowLine(58.0,60.0)(79.0,50.0) 
\Text(36.0,50.0)[r]{$Z$}
\DashLine(37.0,60.0)(37.0,40.0){3.0} 
\Text(15.0,40.0)[r]{$e$}
\ArrowLine(16.0,40.0)(37.0,40.0) 
\Line(37.0,40.0)(58.0,40.0) 
\Text(80.0,30.0)[l]{$e$}
\ArrowLine(58.0,40.0)(79.0,30.0) 
\Text(47,0)[b] {diagr.19}
\end{picture} \ 
{} \qquad\allowbreak
%  diagram # 16
\begin{picture}(95,99)(0,0)
\Text(15.0,80.0)[r]{$e$}
\ArrowLine(16.0,80.0)(37.0,80.0) 
\Line(37.0,80.0)(58.0,80.0) 
\Text(80.0,90.0)[l]{$e$}
\ArrowLine(58.0,80.0)(79.0,90.0) 
\Text(36.0,70.0)[r]{$Z$}
\DashLine(37.0,80.0)(37.0,60.0){3.0} 
\Line(37.0,60.0)(58.0,60.0) 
\Text(80.0,70.0)[l]{$\nu_e$}
\ArrowLine(58.0,60.0)(79.0,70.0) 
\Text(33.0,50.0)[r]{$\nu_e$}
\ArrowLine(37.0,40.0)(37.0,60.0) 
\Text(15.0,40.0)[r]{$e$}
\ArrowLine(16.0,40.0)(37.0,40.0) 
\Text(47.0,44.0)[b]{$W^+$}
\DashArrowLine(58.0,40.0)(37.0,40.0){3.0} 
\Text(80.0,50.0)[l]{$b$}
\ArrowLine(58.0,40.0)(79.0,50.0) 
\Text(80.0,30.0)[l]{$\bar{t}$}
\ArrowLine(79.0,30.0)(58.0,40.0) 
\Text(47,0)[b] {diagr.20}
\end{picture} \ 
}
%\end{document}
\end{center}
\caption{Diagrams for the process $e^- e^- \to e^- \nu_e \bar t
b$, unpolarized beams}
\end{figure}
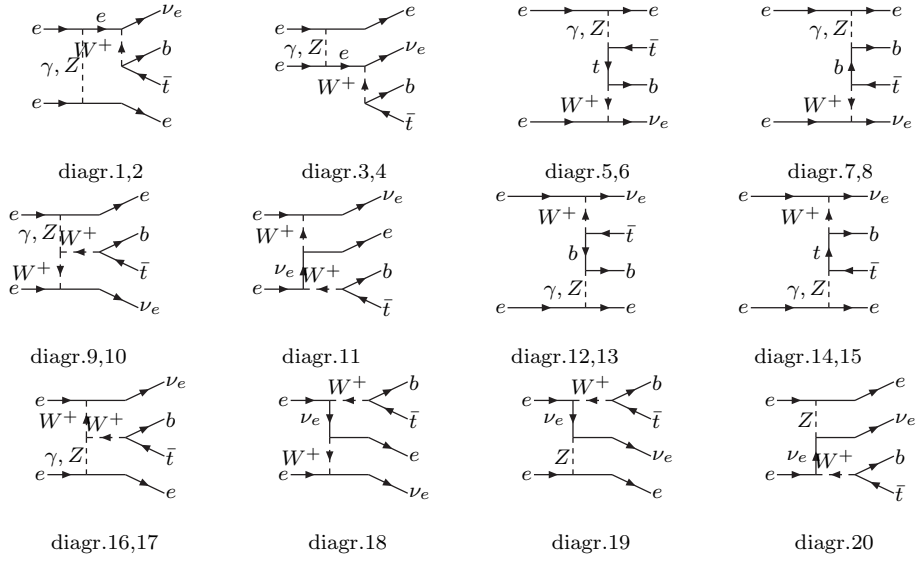
% Fig 4

It is interesting to compare the case of single top production at LC
with the case of single $W$ production at LEP2.
If one replaces the top and $b$ quarks in Figs.1,2
by light quarks, ($t$,$b$ $\to$ $u$, $d$), one gets exactly
the $CC20$ diagrams which were used to study single $W$
production at LEP2 \cite{lep2mc}.
Since the $t \bar t$ cross section at a linear collider is
much larger than the single top cross section (for $\sqrt{s} \geq$350 
GeV), we anticipate a situation similar to the case of single 
$W$ production when the $W$ signal has to be isolated from the large
$W^+ W^-$ pair background. Single
$W$ isolation at LEP2 can be achieved by simple cuts on the forward 
going electron which separate the $t$-channel diagram subset and
minimizes interferences, as concluded from rather extensive
analyses \cite{lep2mc}. However, the LEP2 single $W$ isolation 
procedure cannot be simply repeated in our case. The single top
cross section shown in Fig.7
for unpolarized $e^+ e^-$ collisions is larger than
those of the $t$-channel subset (see Fig.6a), especially at
lower energies, due to considerable single top contributions from  
the $s$-channel subset (Fig.1).
Thus, the separation of events from only the $t$-channel gauge invariant
subset by cuts on forward electron as done at LEP2 
would significantly underestimate the single top cross section.
This rate is underestimated
by e.g. more than
60\% at $\sqrt{s}=$0.5 TeV and by about 30\% at $\sqrt{s}=$1 TeV,
in comparison with the rates obtained from the invariant mass 
subtraction procedure.
The reason can be qualitatively understood as follows. 
The application of cuts on forward electron dramatically 
suppresses the $CC10$ $s$-channel diagrams in both the $W$
and top cases. Since however the ratio of single $W$ to 
$W^+ W^-$ pair cross sections in the $s$-channel subset
is significantly smaller
than the corresponding single top to $t\bar t$ ratio, the
relative contribution of the $CC10$ $s$-channel top diagrams
is enhanced.

\begin{figure}[h!]
\begin{center}
% diagrams for process A,e1 -> n1,T,b             
%\documentstyle[axodraw]{article}
%\begin{document}
{
\unitlength=0.75 pt
\SetScale{0.75}
\SetWidth{0.7}      % line    size control
\scriptsize    %  letter  size control
{} \qquad\allowbreak
%  diagram # 1
\begin{picture}(95,79)(0,0)
\Text(15.0,70.0)[r]{$\gamma$}
\DashLine(16.0,70.0)(37.0,60.0){3.0} 
\Text(15.0,50.0)[r]{$e$}
\ArrowLine(16.0,50.0)(37.0,60.0) 
\Text(47.0,64.0)[b]{$e$}
\ArrowLine(37.0,60.0)(58.0,60.0) 
\Text(80.0,70.0)[l]{$\nu_e$}
\ArrowLine(58.0,60.0)(79.0,70.0) 
\Text(54.0,50.0)[r]{$W^+$}
\DashArrowLine(58.0,40.0)(58.0,60.0){3.0} 
\Text(80.0,50.0)[l]{$b$}
\ArrowLine(58.0,40.0)(79.0,50.0) 
\Text(80.0,30.0)[l]{$\bar{t}$}
\ArrowLine(79.0,30.0)(58.0,40.0) 
\Text(47,0)[b] {diagr.1}
\end{picture} \ 
{} \qquad\allowbreak
%  diagram # 2
\begin{picture}(95,79)(0,0)
\Text(15.0,70.0)[r]{$\gamma$}
\DashLine(16.0,70.0)(58.0,70.0){3.0} 
\Text(80.0,70.0)[l]{$\bar{t}$}
\ArrowLine(79.0,70.0)(58.0,70.0) 
\Text(54.0,60.0)[r]{$t$}
\ArrowLine(58.0,70.0)(58.0,50.0) 
\Text(80.0,50.0)[l]{$b$}
\ArrowLine(58.0,50.0)(79.0,50.0) 
\Text(54.0,40.0)[r]{$W^+$}
\DashArrowLine(58.0,50.0)(58.0,30.0){3.0} 
\Text(15.0,30.0)[r]{$e$}
\ArrowLine(16.0,30.0)(58.0,30.0) 
\Text(80.0,30.0)[l]{$\nu_e$}
\ArrowLine(58.0,30.0)(79.0,30.0) 
\Text(47,0)[b] {diagr.2}
\end{picture} \ 
{} \qquad\allowbreak
%  diagram # 3
\begin{picture}(95,79)(0,0)
\Text(15.0,70.0)[r]{$\gamma$}
\DashLine(16.0,70.0)(58.0,70.0){3.0} 
\Text(80.0,70.0)[l]{$b$}
\ArrowLine(58.0,70.0)(79.0,70.0) 
\Text(54.0,60.0)[r]{$b$}
\ArrowLine(58.0,50.0)(58.0,70.0) 
\Text(80.0,50.0)[l]{$\bar{t}$}
\ArrowLine(79.0,50.0)(58.0,50.0) 
\Text(54.0,40.0)[r]{$W^+$}
\DashArrowLine(58.0,50.0)(58.0,30.0){3.0} 
\Text(15.0,30.0)[r]{$e$}
\ArrowLine(16.0,30.0)(58.0,30.0) 
\Text(80.0,30.0)[l]{$\nu_e$}
\ArrowLine(58.0,30.0)(79.0,30.0) 
\Text(47,0)[b] {diagr.3}
\end{picture} \ 
{} \qquad\allowbreak
%  diagram # 4
\begin{picture}(95,79)(0,0)
\Text(15.0,60.0)[r]{$\gamma$}
\DashLine(16.0,60.0)(37.0,60.0){3.0} 
\Text(47.0,64.0)[b]{$W^+$}
\DashArrowLine(58.0,60.0)(37.0,60.0){3.0} 
\Text(80.0,70.0)[l]{$b$}
\ArrowLine(58.0,60.0)(79.0,70.0) 
\Text(80.0,50.0)[l]{$\bar{t}$}
\ArrowLine(79.0,50.0)(58.0,60.0) 
\Text(33.0,50.0)[r]{$W^+$}
\DashArrowLine(37.0,60.0)(37.0,40.0){3.0} 
\Text(15.0,40.0)[r]{$e$}
\ArrowLine(16.0,40.0)(37.0,40.0) 
\Line(37.0,40.0)(58.0,40.0) 
\Text(80.0,30.0)[l]{$\nu_e$}
\ArrowLine(58.0,40.0)(79.0,30.0) 
\Text(47,0)[b] {diagr.4}
\end{picture} \ 
}
%\end{document}
\end{center}
\caption{Diagrams for the process $\gamma e \to \nu_e \bar t b$}
\end{figure}
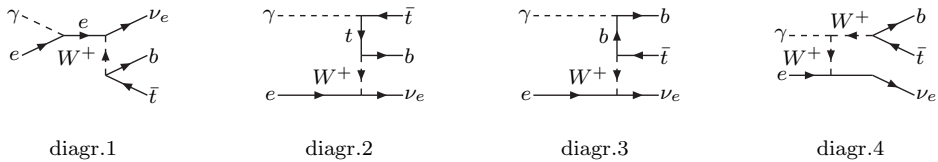
% Fig 5

The availability of longitudinally polarized beams at a linear
collider opens a new window for single top quark studies. 

In the case
of left-handed $e^-$ with right-handed $e^+$ collisions, $e^-_L e^+_R$,
all 20 diagrams of Figs.1,2 contribute to the total top quark
production with a rate three times higher
than for the unpolarized case, see Fig.6b, for the entire reaction
$e^+ e^- \to e^- \bar \nu_e t \bar b$. The 
single top event rate obtained from the mass cut subtraction procedure
is also increased by a factor of three (Fig.7), 
and the energy behaviour resembles, as expected,
the unpolarized cross section increase, reaching 17 fb at 1 TeV.

If right-handed electrons collide with left-handed positrons, $e^-_R 
e^+_L$, only 11 diagrams of Figs.1 and 2 survive. They form 
two gauge invariant subsets: the $t$-channel subset of two diagrams
(diagrams 1 and 2 in Fig.2), and the $s$-channel subset of 9 diagrams 
(diagrams 1-8 and 10 in Fig.1). The single top cross section 
energy behavior,
also shown in Fig.7,
reveals, after a broad maximum around 500 GeV, a slow decrease with 
increasing energy. This happens because the important multiperipheral 
diagrams 
3-8 in Fig.2, responsible for the increase of the
event rate, are removed by orthogonal helicity projectors in the initial 
left-handed positron state and the $W$ boson $(V-A)$ interaction vertex.

If both beam particles collide with right-handed polarizations, 
$e^-_R e^+_R$, only a gauge invariant subset of 9 diagrams 
contribute (all diagrams in Fig.2 
except diagram 9). Hence, this helicity 
configuration
avoids a priori the $t \bar t$ background and would 
provide a good laboratory for single top quark physics.
Cross 
sections are large enough for precise measurements, see Fig.7, 
being close to 1 fb
at $\sqrt{s}=$0.5 TeV respectively 8 fb at $\sqrt{s}=$1 TeV.

For completeness it is worth mentioning that $e^-_L e^+_L$ collisions 
involve only two diagrams (diagrams 1,2 in Fig.2) with negligibly 
small cross sections due to unfavorable helicity configurations.
They are formed by means of the left-handed projector of the initial 
positron and the left-handed $e^-\nu_e$ vector current. 
\begin{figure}[h!]
\begin{picture}(11,11)
\put(1.3,-1.5){\epsfxsize=10cm
         \epsfysize=13 cm \leavevmode \epsfbox{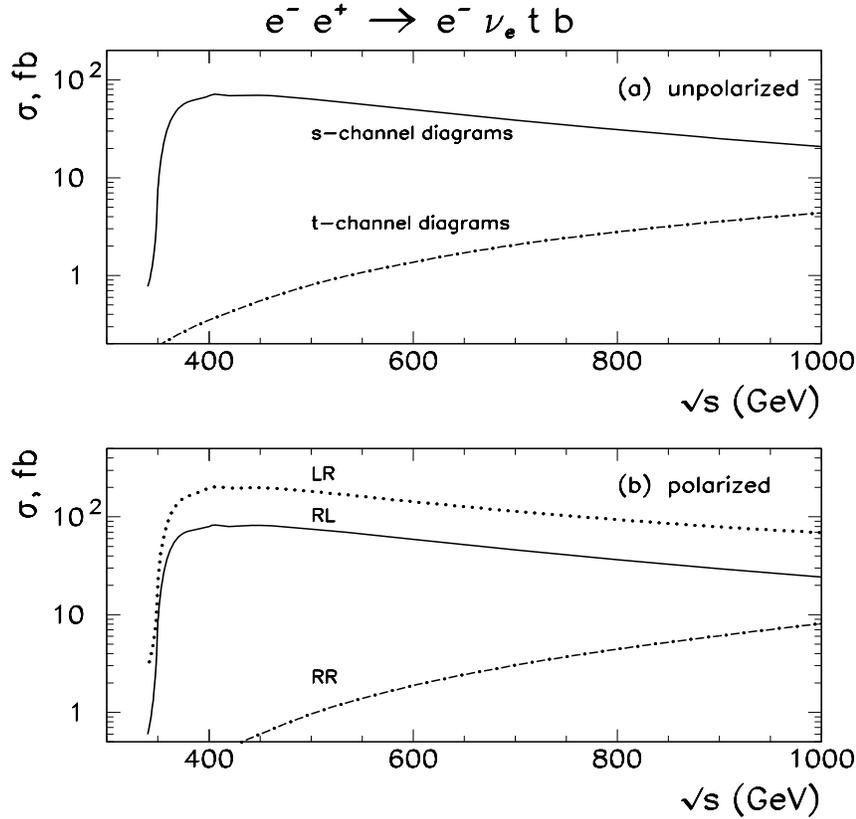}}
\end{picture}
\caption{Energy dependence of the cross sections for the process $e^+
e^-\to e^- \bar \nu_e
t \bar b$ for (a) unpolarized $e^+ e^-$ $s$- and $t$-channel
gauge invariant subsets and (b) fully polarized
beams}
\end{figure}
% Fig 6

It is important to point out that for both unpolarized and
opposite-polarized $e^+ e^-$ collisions the dominating 
$t \bar t$ event rates have to be subtracted from 
the total rate, resulting in additional statistical
uncertainties for single top cross section measurements.
For example, in the unpolarized $e^+ e^-$ case at 
$\sqrt{s}=$500 GeV and an integrated
luminosity of 500 fb$^{-1}$ the relative single top cross section
error $\delta \sigma/\sigma$ increases by 8.2\%.

In summary, in $e^+ e^-$ collisions
single top quark production is largest for the $e^-_L e^+_R$ case,
but its experimental precision is diluted by additional statistical 
uncertainties from the $t \bar t$ subtraction procedure. $e^-_R e^+_R$
collisions avoid a priori the $t \bar t$ background and yield, in 
particular
at large $\sqrt{s}$, cross sections of a few fb, which is sufficient 
for precise measurements at a high-luminosity collider. 
However, this statement should be somewhat qualified due to 
partial beam 
polarizations in real experiments. For collisions of e.g. 80\% 
right-polarized electrons with 60\% right-polarized positrons we expect 
significant $t \bar t$ contaminations from $e^-_R e^+_L$ and $e^-_L e^+_R$ 
initial states of 44.1 (14.9) fb at 0.5 (1.0) TeV. These values should be 
compared with the expected single top cross sections of 8.6 and 12.1 fb,
respectively, so that due to imperfect beam polarizations  
the $e^-_R e^+_R$ configuration turns out to be very similar
to the unpolarized, right-left or left-right configurations at least 
at $\sqrt{s}=$0.5 TeV.
\begin{figure}[h!]
\begin{picture}(10.5,10.5)
\put(1.8,-1.8){\epsfxsize=10cm
         \epsfysize=13 cm \leavevmode
          \epsfbox{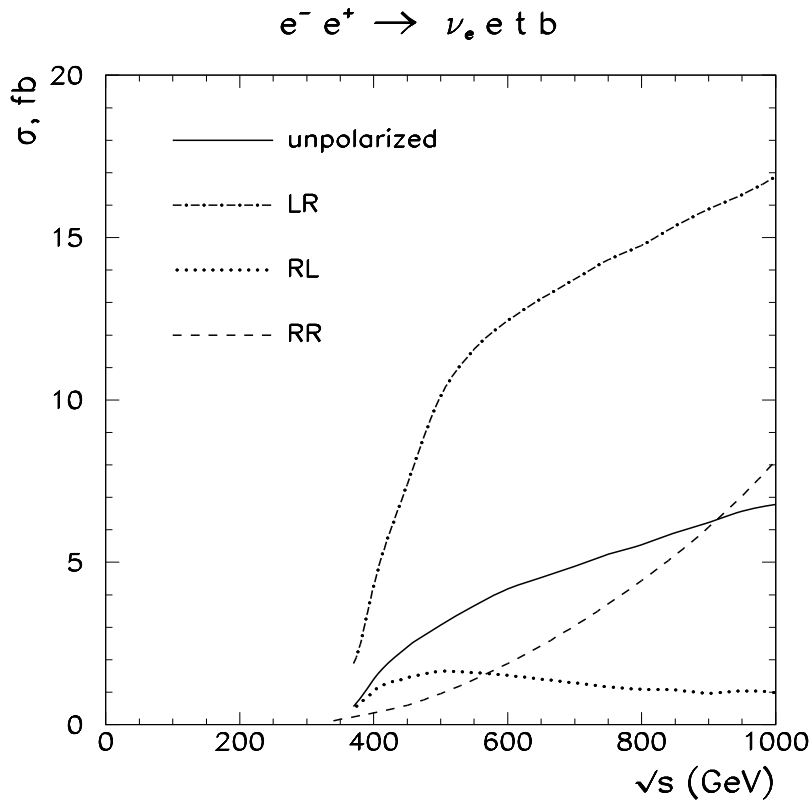}}
\end{picture}
\caption{Energy dependence of the cross section for single top quark 
production in the process
$e^+ e^-\to e^- \bar \nu_e t \bar b$, for unpolarised and fully
polarized beams.}
\end{figure}
% Fig 7
\subsection{Single top production in $e^- e^-$ collisions with
unpolarized and polarized beams}

Due to charge conservation no $s$-channel $\gamma$, $Z$ diagrams
are possible in $e^- e^-$ collisions. Consequently, in the reaction
$e^- e^- \to e^- \nu_e \bar t b$ the top quark is only singly produced.
If both electrons are unpolarized, 20 diagrams shown in Fig.4 contribute.
The energy dependence of the cross section is shown in Fig.8. It
grows from almost 2 fb at 0.5 TeV and to 19 fb
at 1.0 TeV. Thus, at a high-luminosity $e^- e^-$ collider precise single 
top cross section measurements can be performed, with the advantage of no 
need to subtract large $t \bar t$ background. 
\begin{figure}[h!]
\begin{picture}(10,10.5)
\put(1.8,-1.8){\epsfxsize=11cm
         \epsfysize=13 cm \leavevmode \epsfbox{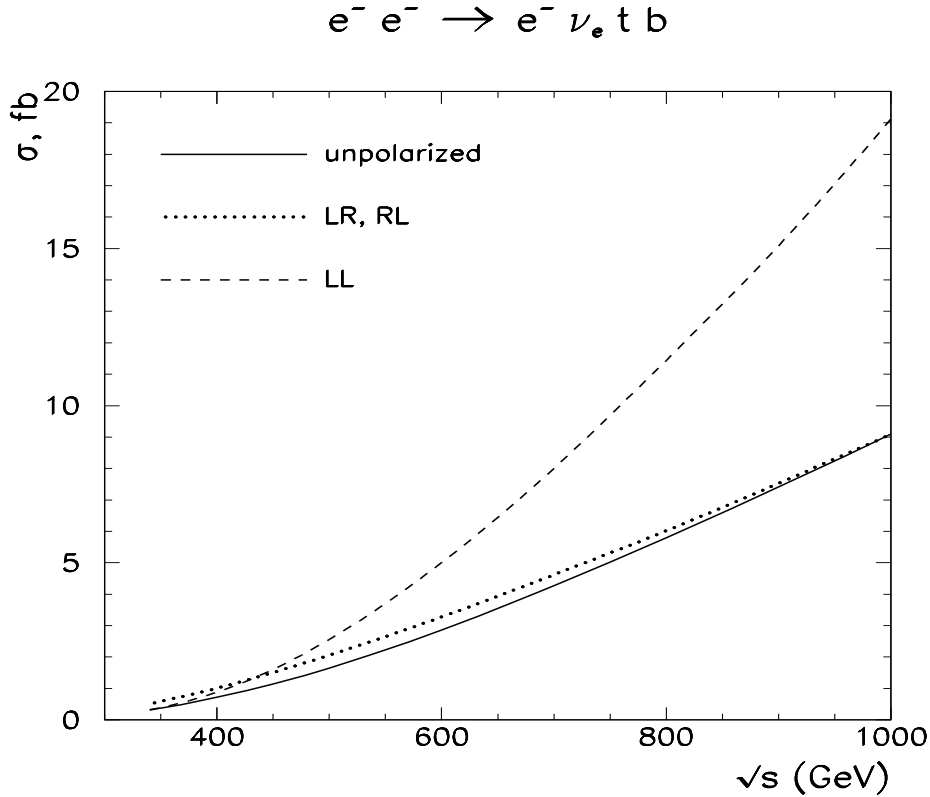}}
\end{picture}
\caption{Energy dependence of the cross section for the processes   
$e^- e^- \to e^- \nu_e \bar t b$,
for unpolarized and fully polarized beams}
\end{figure}
% Fig 9

The cross section for collisions of
left-handed and right-handed electrons, $e^-_L e^-_R$, also 
shown in Fig.8, is very close to the unpolarised case.
Here only 11 diagrams (1-10 and 19 in Fig.4) out of the 20 
survive.

If both electrons are left-handed polarized, $e^-_L e^-_L$, all 
20 diagrams of Fig.4 contribute as in the unpolarized case, but 
the cross section is larger by a factor of two. Since $e^-$ beams can be
easily polarized to a high degree (more than 80\% at SLAC routinely
and may well be raised to 90\% by the time the LC is built), 
$e^-_L e^-_L$ scattering is very well suitable for single top cross 
section measurements with high precision. Right-polarized electron
collisions,
$e^-_R e^-_R$, with only yields from diagrams 1-4 in Fig.4, 
give however a cross section 50 times smaller than $e^-_L e^-_L$
collisions (see Table 1).

We would like to mention that $e^+ e^-$ and $e^- e^-$ single top 
event rates are comparable. In the $e^+ e^-$ case, significant 
$s$- and $t$-channel contributions were found, contrary 
to the $e^- e^-$ case where only $t$-channel diagrams exist. One
might expect that 
$e^- e^-$ collisions are more sensitive to 
effects of new physics (e.g. to anomalous $Wtb$ couplings), insofar as,
in general, $t$-channel topologies have larger sensitivity to
nonstandard vertices than the $s$-channel ones.   
\begin{figure}[h!]
\begin{picture}(10.5,10.5)
\put(1.8,-1.8){\epsfxsize=11cm
         \epsfysize=13 cm \leavevmode
          \epsfbox{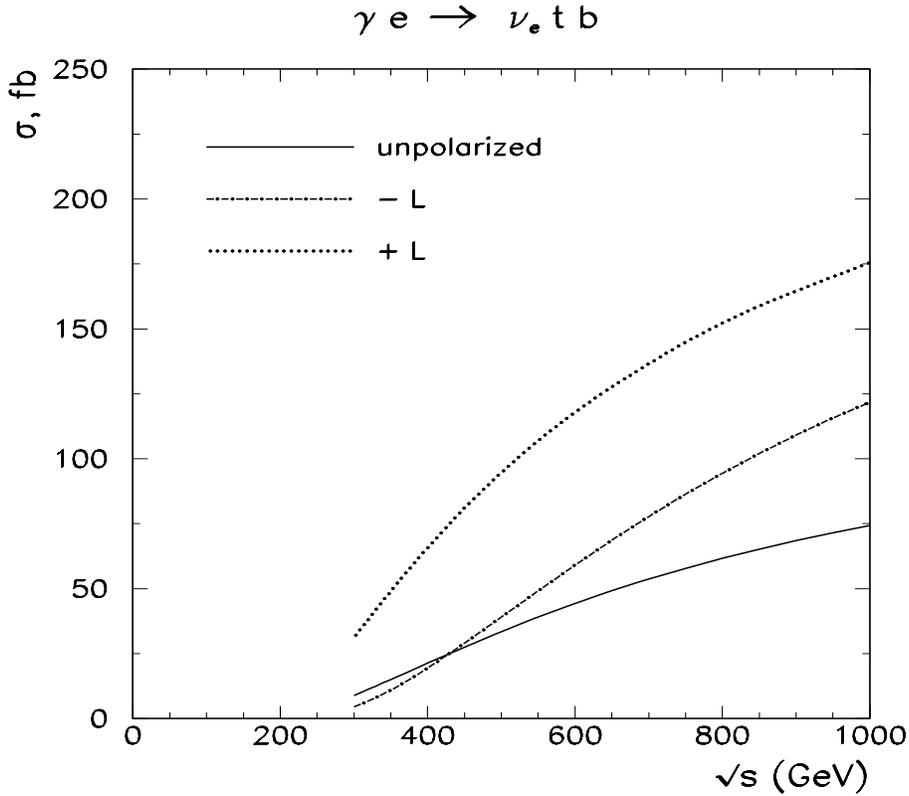}}
\end{picture}
\caption{Energy dependence of the cross section for the process
$\gamma e^-\to \nu_e \bar t b$, for unpolarized and fully
polarized beams.}
\end{figure}
% Fig 12

In the SM the $Wtb$ coupling is proportional to 
the CKM matrix element $V_{tb}$. Assuming 100 fb$^{-1}$ 
integrated
luminosity and full reconstruction efficiency, about 200 events 
expected at $\sqrt{s}=$500 GeV would allow one to measure  
$V_{tb}$ with an uncertainty
of 7\% at the $2\sigma$ level. This accuracy is of
the same order as expected from Tevatron and LHC experiments 
\cite{beneke}.

\subsection{Single top production in $\gamma e$ collisions with
unpolarized and polarized beams}

The  $\gamma e$ mode of a linear collider offers unique properties
in favor of single top quark physics in the reaction
$\gamma e \to \nu_e \bar t b$. No $t \bar t$ background exists,
the cross section for unpolarized collisions is large
\cite{ge_unpol,ge}, high degree of beam polarizations is
possible and the number of contributing diagrams (Fig.5) is
only four. For these reasons, the potential sensitivity to the $Wtb$ 
coupling is expected to be high. In this paper we extend
previous studies \cite{ge_unpol,ge} to collisions of polarized
electrons with either unpolarized or polarized ($+,-$)
photons. Cross sections for unpolarized beams and 
polarized photons colliding with left-handed 
electrons are shown in Fig.9. If electrons are unpolarized,
the corresponding polarized cross sections must be divided by two.
The most favoured case of $\gamma_+ e^-_L$ collisions results to
approximately 100 (180) fb at 0.5 (1.0) TeV. Cross sections with
right-handed electrons are suppressed by a factor of the order 
$m^2_e/s$ and thus negligibly small (unfavorable helicity configuration).
\begin{figure}[h!]
\begin{picture}(10.5,10.5)
\put(1.8,-1.8){\epsfxsize=11cm
         \epsfysize=13 cm \leavevmode
          \epsfbox{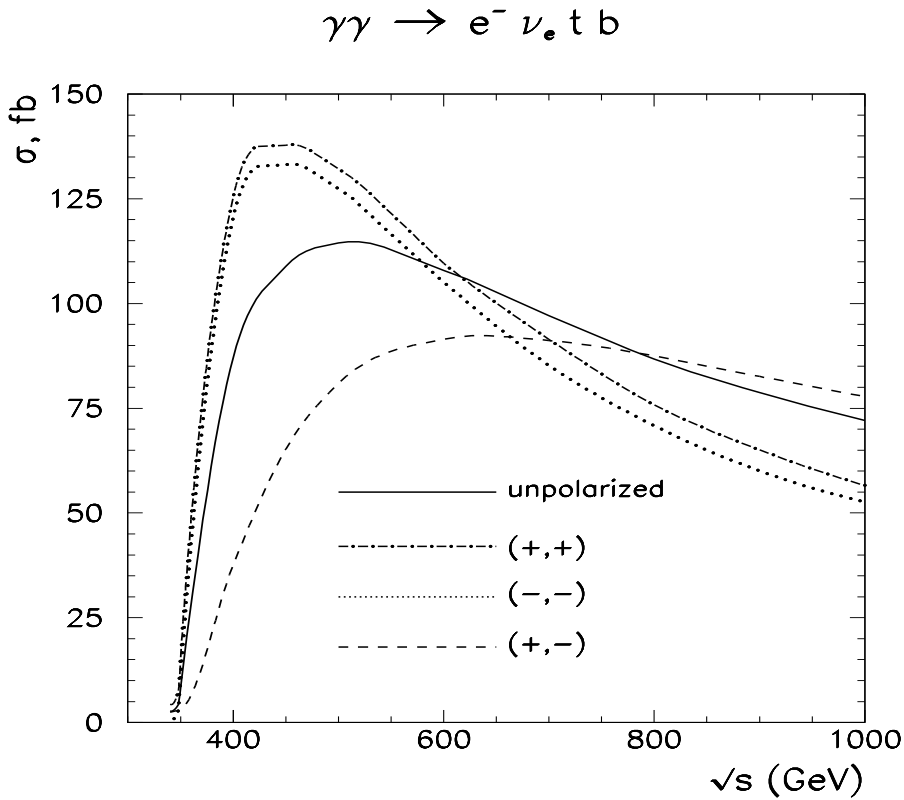}}
\end{picture}
\caption{Energy dependence of the cross section for
the process $\gamma \gamma \to e^- \bar \nu_e t \bar b$,   
for unpolarized and fully polarized beams.}
\end{figure}

With 10$^4$ events from $\gamma_+ e^-_L$ collisions at
$\sqrt{s}=$0.5 TeV,
expected for 100 fb$^{-1}$ integrated luminosity
and no event loss, we measure  $V_{tb}$ with an uncertainty
of 1\% at the $2\sigma$ level. This excellent precision can be
achieved neither at the Tevatron nor the LHC and from a LC 
top quark width measurement at the $t \bar t$ threshold, 
with anticipated uncertainties of approximately 5\% \cite{beneke} 
respectively 10\% \cite{comas}.
In summary, the reaction $\gamma e \to \nu_e \bar t b$ with unpolarized or
properly polarized beams is extremely suitable for precise measurements of
important top quark properties.
\begin{figure}[h!]
\begin{picture}(10.5,10.5)
\put(1.8,-1.8){\epsfxsize=11cm
         \epsfysize=13 cm \leavevmode
          \epsfbox{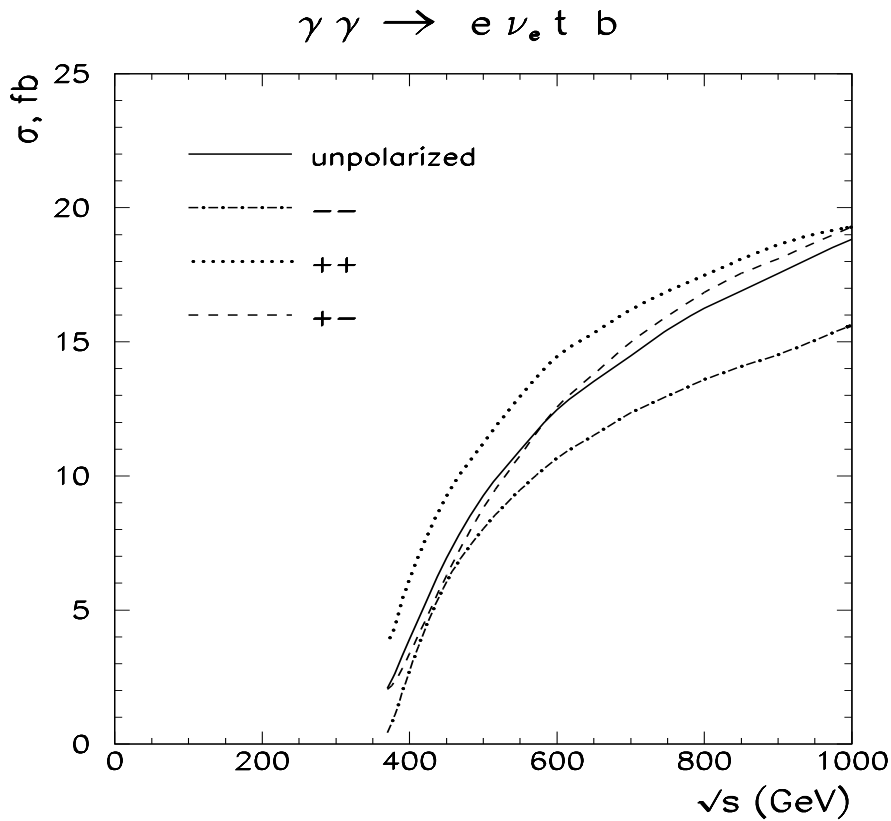}}
\end{picture}
\caption{Energy dependence of the cross section for single top production
in the process
$\gamma \gamma \to e^- \bar \nu_e t \bar b$,
for unpolarized and fully polarized beams.}
\end{figure}
% Fig 11
\subsection{Single top production in $\gamma \gamma$ collisions
with unpolarized and polarized beams}

Additional possibilities of single top quark production are offered
by the $\gamma \gamma$ mode of a linear collider. The reaction
$\gamma \gamma \to e^- \bar \nu_e t \bar b$ however is dominated by
$t \bar t$ pair production for all initial polarization 
states possible,
and requires always $t \bar t$ subtraction to obtain the single top 
production rate. All contributing diagrams are shown in Fig.3.
Unlike the $e^+ e^-$ case, there are no gauge invariant subsets
within the complete set of 21 diagrams.
The total cross sections of the 
reaction
$\gamma \gamma \to e^- \bar \nu_e t \bar b$ for unpolarized
and polarized beams are shown in Fig.10, and
the $t \bar t$ subtracted single top quark production rates
in Fig.11. Obviously, a preferred photon helicity configuration is not
evident from Fig.11. In comparison to $e^+ e^-$ and $e^- e^-$ collisions,
single top cross sections in $\gamma \gamma$ collisions are enhanced,
in particular at energies around 500 GeV. However, this advantage 
is expected to be degraded by the 
lower luminosity of a Compton collider. 

It is worthwhile to compare single top production in 
gamma-gamma collisions at LC with single top production 
in gluon-gluon partonic subprocess at the LHC \cite{LHC00,LHC1}
since they yield the same final state $e^- \nu_e t \bar b$. 
The $t \bar t$ pair production part in $gg$ 
collisions is removed by means of the same gauge invariant subtraction 
procedure as described in section 2.1. However, another problem 
in the case of $gg$ collisions is the double counting in the sum of the 
$gg \to tWb$
and the $gb\to tW$ partonic subprocesses, and a gauge invariant
subtraction of the $g \to b \bar b$ splitting term is necessary to avoid 
it. One might expect that a similar double counting 
problem arises in the case of $\gamma \gamma$ collisions if one
includes the 'resolved photon' contribution, when the $b$
constituent of the first 'resolved photon' collides with the second photon. 
The yield of 'resolved photon' (with poorly known gluon and $b$ contents) 
is however expected to be very small at the characteristic momentum
transfer scale $Q^2 \sim m^2_b$. So, the problem
of double counting in $\gamma \gamma$ collisions is expected 
to be not important compared to $gg$ collisions.

\section{Single top production with anomalous \\
effective operators of the third generation quarks}

Single top quark production at linear colliders is thought
to be a promising tool to probe the $Wtb$ vertex.
Simple counting of single top events is sensitive
to anomalous $Wtb$ couplings, unlike
the $t\bar t$ pair production where 
deviations from the SM are difficult to notice.
A sophisticated combination of e.g.
forward-backward, spin-spin and energy asymmetries is needed to probe
couplings in the $Wtb$ vertex (\cite{BDSS}, see also \cite{recent}).

In the following we demonstrate the sensitivity to anomalous
$Wtb$ couplings for the single top quark reactions
$e^+_R e^-_R \to e^- \bar \nu_e t \bar b$,
$e^-_L e^-_R \to e^- \nu_e \bar t b$ and
$\gamma_+ e^-_L \to \nu_e \bar t b$, which, as outlined in sect.2,
are very promising for this task. Their amplitudes are directly
proportional to the $Wtb$ coupling and $t \bar t$ background is absent.

In order to probe anomalous $Wtb$ couplings in a model independent
way, we use the effective Lagrangian of dimension 6 as proposed in
\cite{KLY}
\begin{eqnarray}
{\cal L}  =  \frac{g}{\sqrt{2}}
% [ W_{\mu}^-\bar{b} \gamma_{\mu} P_L t
    \frac{1}{2m_W} W_{\mu\nu}
\bar{t}\sigma^{\mu\nu}(f_{2R} P_L + f_{2L} P_R) b + {\rm h.c.}
\end{eqnarray}
Here $f_{2L}$ and $f_{2R}$ are the anomalous couplings, 
$W_{\mu\nu} =
D_{\mu}W_{\nu} - D_{\nu}W_{\mu}, D_{\mu} = \partial_{\mu} - i e A_{\mu}$,
$P_{L,R} = 1/2(1 \pm \gamma_5)$ and
$\sigma_{\mu \nu}=i(\gamma_{\mu} \gamma_{\nu}-\gamma_{\nu} 
\gamma_{\mu})$/2.

Calculations of diagrams with photons in the $t$-channel
using the effective $Wtb$ vertex (3) 
should be carefully performed. It follows from existing
experience that the direct introduction of Breit-Wigner propagators with
finite width in the amplitude
could break the unitary behaviour of the single top 
cross section. Unitary behaviour is ensured by the cancellation
of the double pole $1/t^2$ behaviour of the 
individual photon exchange squared diagrams to the single pole $1/t$ 
behaviour of the squared amplitude 
\footnote{ It is important to point out that
additional 4-point vertices $\gamma W t b$ appear due to
covariant derivatives in the $W^{\pm}_{\mu \nu}$ tensor (see 
\cite{ge_unpol} for details)}.
This 
cancellation is controlled by the electromagnetic $U(1)$ gauge invariance. 
The violation of the $U(1)$ invariance is usually manifested as
a several orders of magnitude cross section increase
\cite{shimizu,oldenborgh} for
very small electron scattering angles in obvious contradiction with
the unitary cross section behaviour.
The violation of unitary 
behaviour is well known from 
analyses of single $W$ production in the SM \cite{shimizu} and various 
prescriptions to circumvent this difficulty were proposed
(see for instance \cite{singleW} and references therein).
We are using the process-independent {\it overall} prescription for
Breit-Wigner propagators (for details see \cite{Bardin,shimizu,singleW}).

The cancellation of the
$1/t^2$ pole can be checked directly by inspection of the differential
cross section $d\sigma / d{\tt log}(t)=t\, d\sigma / dt$ 
\cite{singleW,oldenborgh}. As an example, this cross section
is shown in Fig.12
for the process $e^-_R e^-_L \to e^- \nu_e \bar t b$.
At small $|t|$ near the pole,
the distribution for the anomalous
couplings $f_{2L}=$-0.5, $f_{2R}=$0 is only marginally 
different
in comparison with the SM cross section ($f_{2L}=f_{2R}=$0), so
the cancellation of the double pole indeed occurs.
\begin{figure}[h!]
\begin{picture}(10.5,10.5)
\put(1.8,-1.8){\epsfxsize=11cm
         \epsfysize=13 cm \leavevmode
          \epsfbox{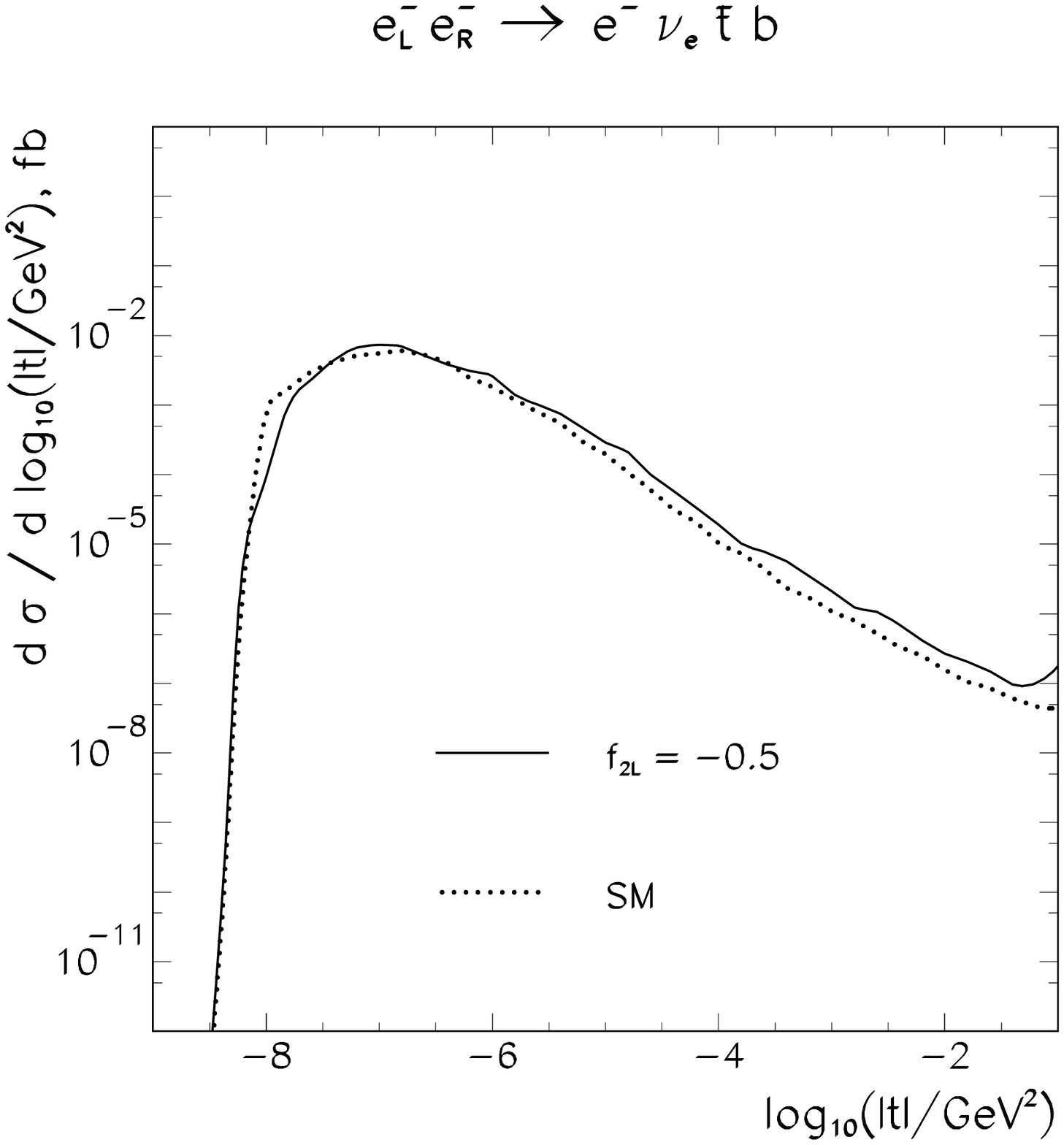}}
\end{picture}
\caption{Distributions in the logarithm of the $t$-channel photon momentum
transfer squared for the process $e^-_L e^-_R \to e^- \nu_e \bar t b$
in the SM and for the anomalous couplings $f_{2L}=$-0.5 and $f_{2R}=$0}
\end{figure}
% Fig 13

Calculations with different values for $f_{2L}$ and $f_{2R}$
were performed for the three reactions mentioned above, mostly
at $\sqrt{s}=$500 GeV and for two values of integrated luminosity.
The 2$\sigma$ bounds of the anomalous coupling parameter space,
within which no distinction from the SM is possible,
are shown in Fig.13.
Best bounds for $f_{2L}$ can be obtained from the reaction
$\gamma_+ e^-_L \to \nu_e \bar t b$, \\
-0.02 $\leq$ $f_{2L}$ $\leq$ 0.06,
being relatively independent of the integrated luminosity. In comparison 
to unpolarized $\gamma e^-$ collisions \cite{ge_unpol}, the beam
polarizations 
restrict the allowed range of $f_{2L}$ by a factor of 2 to 3,
but have no impact on $f_{2R}$.
The coupling $f_{2R}$ is almost equally well bounded to
-0.1 $\leq$ $f_{2R}$ $\leq$ 0.1 by $e^-_R e^+_R$ and $\gamma_+ e^-_L$
collisions
at 500 GeV for 100 fb$^{-1}$ integrated luminosity. Improvements for
$f_{2R}$ can be achieved by e.g. a five times higher luminosity
in either the $e^-_L e^-_R$ or the $e^-_R e^+_R$ case, with best
2$\sigma$ bounds of  -0.05 $\leq$ $f_{2R}$ $\leq$ 0.05 from
the reaction
$e^-_R e^+_R \to e^- \bar \nu_e t \bar b$. Here, doubling the c.m.s.
energy affects $f_{2R}$ in the same way as an increase
of the integrated luminosity from 
100 fb$^{-1}$ to 500 fb$^{-1}$.

The introduction of effective operators with the top quark
is usually motivated by the expectations that the SM is an
effective theory at the electroweak scale for some underlying
new physics at higher energies. Anomalous interactions of the third
generation fermions ($t$,$b$) are introduced by most
general local $SU(3)\times
SU(2)\times U(1)$ invariant effective Lagrangian terms
including also Higgs and gauge bosons. In principle,  
various effective operators can be constructed. A recent discussion
of operators with dimension 4 and 5 can be found in \cite{tait}.
They are strongly constrained by the experimental data
on $b$-quark decays and do not contribute to top quark
cross sections at a level sufficient for experimental observation.  
However, for the $CP$ and flavor conserving effective operators of 
dimension 6 the limits from experimental data are weaker or
even missing. In the notation of \cite{whisnant}
seven effective operators ($SU(3)\times SU(2)\times U(1)$ invariant
before the electroweak symmetry breaking) provide the anomalous
contributions to the $Wtb$ vertex:
$O_{qW}$, $O^3_{\Phi q}$, $O_{Db}$, $O_{bW \Phi }$,
$O_{Dt}$, $O_{tW \Phi }$, $O_{t3}$. 
Not all of them are equally important for linear
collider studies.
The operators $O_{qW}$ and $O^3_{\Phi q}$ are severely constrained
by
the LEP1 observables $R^b$ and $A^b_{FB}$.
The remaining operators are so far not directly constrained by any 
experimental data. Bounds from partial wave 
unitarity arguments can be found in \cite{gounaris}. 
The operator $O_{t3}$ has
the same structure as dimension 4 right-handed current 
and is strongly restricted for reasons as given in
\cite{tait}.
The operators $O_{Dt}$ and $O_{Db}$ include the derivatives of
the $t$ and $b$ fields. If one applies the equations of motion
to them \cite{buchmuller}, they can
be expressed within another class of effective operators. 
\begin{figure}[h!]
\begin{picture}(17,18.5)
\put(0,0){\epsfxsize=14cm
         \epsfysize=16 cm \leavevmode
          \epsfbox{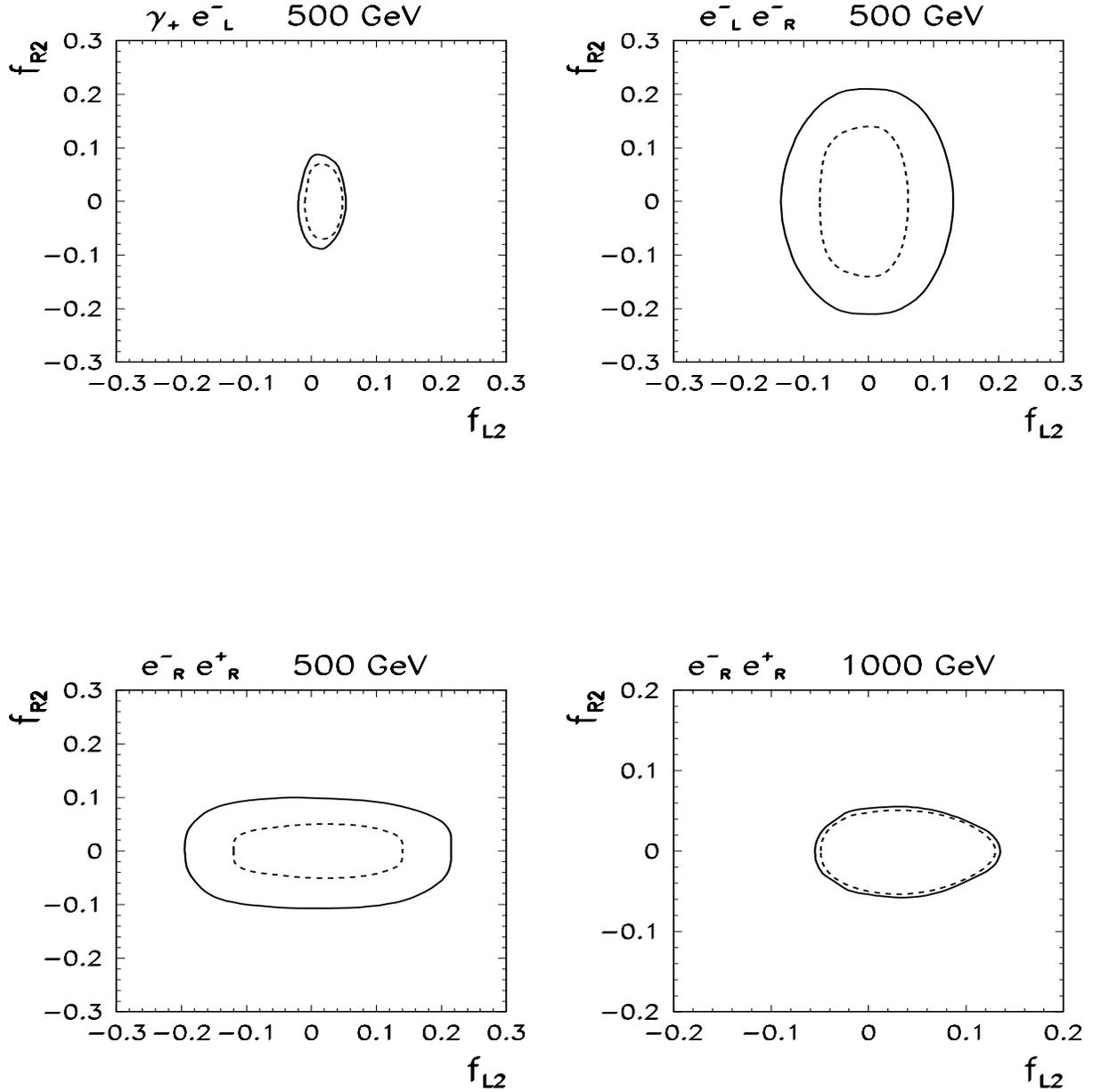}}
\end{picture}
\caption{ 2$\sigma$ bounds on the anomalous couplings $f_{2L}$ and
$f_{2R}$ from the reactions $\gamma_+ e^-_L \to \nu_e \bar t b$,
$e^-_L e^-_R \to e^- \nu_e \bar t b$ and
$e^-_R e^+_R \to e^- \bar \nu_e t \bar b$
at
$\sqrt{s}=$0.5 TeV and 1.0 TeV, for integrated luminosities of 100
fb$^{-1}$ (solid lines) and 500 fb$^{-1}$ (dashed lines).}
\end{figure}
% Fig 14

So only the $O_{bW \Phi }$ and $O_{tW \Phi }$ operators remain:
\vspace{0cm}
\begin{eqnarray}
O_{tW \Phi }& =& [(\bar{q}_L \sigma^{\mu \nu} \tau^I t_R ) \Phi
            + \Phi^{+} (\bar{t}_R \sigma^{\mu \nu} \tau^I q_L)]
            W^I_{\mu \nu} \\
O_{bW \Phi } &=& [(\bar{q}_L \sigma^{\mu \nu} \tau^I b_R ) \Phi
            + \Phi^{+}(\bar{b}_R \sigma^{\mu \nu} \tau^I q_L)]
            W^I_{\mu \nu}
\end{eqnarray}
Here 
$q_L$ is the left-handed third-family
doublet, $\Phi$ is the Higgs boson doublet, $\tau^I=\sigma^I/2$, $W_{\mu}
= \tau^I W^I_{\mu}$.
In the unitary gauge the following effective
$Z b\bar b$, $\gamma b \bar b$, $W t \bar b$, $Z t \bar t$ and $\gamma
t \bar t$ Lagrangian terms appear after symmetry breaking and the 
rotation to  
physical fields $(W^3_{\mu}$, $B_{\mu}$) $\to$ ($Z_{\mu}$, $A_{\mu}$)
\begin{eqnarray}
{\cal L}_{W t \bar b} &=&
\frac{C_{tW \Phi }}{\Lambda^2} \frac{v}{2} W^+_{\mu \nu} (\bar t
 \sigma^{\mu \nu} P_L b)+
\frac{C_{bW \Phi }}{\Lambda^2} \frac{v}{2} W^+_{\mu \nu} (\bar t
 \sigma^{\mu \nu} P_R b) \\ 
{\cal L}_{Z b \bar b} &=&
\frac{C_{bW \Phi }}{\Lambda^2} \frac{c_W}{2}\frac{v}{\sqrt{2}}
Z_{\mu \nu} \bar b \sigma^{\mu \nu} b \\ 
{\cal L}_{\gamma b \bar b} &=&
-\, \frac{C_{bW \Phi }}{\Lambda^2} \frac{s_W}{2}\frac{v}{\sqrt{2}}
A_{\mu \nu} \bar b \sigma^{\mu \nu} b \\
{\cal L}_{Z t \bar t} &=&
-\, \frac{C_{tW \Phi }}{\Lambda^2} \frac{c_W}{2}\frac{v}{\sqrt{2}} 
Z_{\mu \nu} \bar t \sigma^{\mu \nu} t \\
{\cal L}_{\gamma t \bar t} &=&
\frac{C_{tW \Phi }}{\Lambda^2} \frac{s_W}{2}\frac{v}{\sqrt{2}}
A_{\mu \nu} \bar t \sigma^{\mu \nu} t 
\end{eqnarray}
where $\Lambda$ is the scale of new physics, $v\, \sqrt{g^2+g'^2}=2m_Z$,
$s^2_W=1-m^2_W/m^2_Z$ and the $C$'s denote the couplings.
The corresponding Lagrangian is only $U(1)$ invariant. So, if we
start from the $SU(2)\times U(1)$ invariant operators (4) and (5),
the introduction of the anomalous couplings $C_{bW \Phi}$
and $C_{tW \Phi}$ in a gauge 
invariant manner inevitably gives
also anomalous contributions to the $Vb\bar b $ and $Vt\bar t$
($V=\gamma,Z$) vertices (6)-(10) involving the same couplings.   
After the redefinition of the anomalous couplings   
$f_{2L}$ and $f_{2R}$ in (3)
\begin{equation}
f_{2L} = \frac{C_{t W \Phi}}{\Lambda^2} \frac{v\sqrt{2} \, m_W}{g}, \quad
f_{2R} = \frac{C_{b W \Phi}}{\Lambda^2} \frac{v\sqrt{2} \, m_W}{g}
\end{equation}
we obtain exactly the effective $Wtb$ term (6). 

Usually, in analyses of anomalous couplings
the lowest order $s$-channel diagrams
with only one
anomalous vertex are calculated in the {\it production
$\times$ decay} approximation.
However, in our case of three and four fermion final states,
gauge invariant subsets of diagrams with all five
effective vertices (6)-(10) exist, and 
$SU(2)$ symmetry conserving
calculations should account for all vertices
(6)-(10) at the 
same time. Such calculations are beyond our present techniques.
Existing experience of SM calculations in approaches where
$SU(2)$ symmetry is violated, result in an increase of cross sections 
by a factor of 2 to 3. This 
discrepancy is not so dramatic as in the case of broken $U(1)$, with 
deviations of several orders of magnitude. 
 
So consistent and rigorous investigation of top quark physics 
induced by $SU(2)\times U(1)$ invariant local operators  
(4) and (5) is, generally speaking, nontrivial and requires careful 
additional studies. It is interesting to analyse the sensitivity
of the helicity suppressed SM processes 
(like $\gamma e^-_R \to \nu_e \bar t b$ or $e^+_L e^-_L \to e^- \bar
\nu_e t \bar b$) to anomalous effective operators which could destroy
such a $m^2_e/s$ suppression and might lead to observable cross sections.

\section{Conclusions}

Top quark physics will be one of the central issues of the
physics program for a next linear collider. In particular, nonstandard
phenomena are expected to be most pronounced in the top quark   
sector \cite{peccei}. As discussed in this paper and elsewhere,   
single top quark production processes
imply distinct advantages in measurements of the $V_{tb}$
matrix element and the structure of the $Wtb$ vertex.

Taking into account the variety of possible collision modes
for the LC
($e^+ e^-$, $e^- e^-$, $\gamma e^-$, $\gamma \gamma$), combined
with all possibilities of different beam polarizations,
we performed single top cross section evaluations
for the complete sets of the SM tree-level diagrams.
From comparisons of the possible production rates 
we conclude that the extremely favoured single top production
process is $\gamma e \to \nu_e \bar t b$, especially in the case
of polarized collisions. Precise 
cross section
measurements are accessible due to large counting rates and the
absense of $t \bar t$ pair production.
The best option in $\gamma e$ collisions is found when circular polarised
($+$) photons collide with left-handed electrons,
at the largest possible energy. In this case,
$\sigma(\gamma_+ e_L \to \nu_e \bar t b)$
is close to 100 fb at 0.5 TeV and grows to about 170 fb at 1 TeV
c.m.s. energy. Thanks to the proportionality of 
$\sigma(\gamma e \to \nu_e \bar t b)$
to the CKM matrix element $V_{tb}$, unrivaled
precision can be achieved for the latter.

The process $e^- e^- \to e^- \nu_e \bar t b$ is also appropriate for
precise single top cross section measurements. Here the clean
environment, the relatively simple switchover mechanism from $e^+ e^-$
to $e^- e^-$ collisions, no $t \bar t$ pair production and large electron
polarization degree combined with cross sections of $\geq$2 fb
make this process promising.

Reactions like $e^+ e^- \to e^- \bar \nu_e t \bar b$ and 
$\gamma \gamma \to e^- \bar \nu_e t \bar b$ are less favoured for single 
top quark physics. The separation of $t \bar t$ production, typically
about two orders of magnitude larger than single top rates, dilutes the 
precision of single
top quark cross section determination. Even if both $e^+$ and $e^-$
are right-polarized, so that no $t \bar t$ production is possible, 
imperfect polarizations degrade significantly a precise
$e^+_R e^-_R \to e^- \bar \nu_e t \bar b$ event rate measurement.

As an illustrative example of the objectives of single top quark 
production processes we analysed their sensitivity to anomalous $Wtb$ 
couplings $f_{2L}$, $f_{2R}$ of the effective dimension 6 Lagrangian.
A large single top cross section 
does not necessarily provide better sensitivity to anomalous couplings.
We selected from all possibilities the following reactions 
$\gamma_+ e^-_L \to \nu_e \bar t b$,
$e^+_R e^-_R \to e^- \bar \nu_e t \bar b$ and
$e^-_L e^-_R \to e^- \nu_e \bar t b$, since they are
relatively simple and free of $t\bar t$ background.
At $\sqrt{s}=$0.5 TeV and  
assuming 100 fb$^{-1}$ or 500 fb$^{-1}$ integrated luminosity,
best bounds for $f_{2L}$
( -0.02 $\leq f_{2L} \leq $0.06 ) were found for the reaction
$\gamma_+ e^-_L \to \nu_e \bar t b$. The coupling $f_{2R}$ is 
expected to be almost equally well bounded ( -0.1$\leq f_{2R} \leq $0.1 )
by $\gamma_+ e^-_L$ and $e^+_R e^-_R$ collisions, 
with the potential for improvements to -0.05 $\leq f_{2R} \leq$ 0.05 in 
the fully polarized $e^+_R e^-_R \to e^- \bar \nu_e t \bar b$ channel
for 500$fb^{-1}$ integrated luminosity.
These bounds are comparable or somewhat improved to those from $t \bar t$ 
studies, where sophisticated combinations of asymmetries \cite{BDSS}
are needed to probe the anomalous $Wtb$ vertex.  

We are aware that our results rely on tree-level calculations and it 
would be desirable to include next-to-leading order and initial state 
radiation corrections, beamstrahlung and to account for a realistic 
backscattered photon spectra. We however do not expect that our 
conclusions will be qualitatively affected by more elaborate calculations. 

\begin{center}
{\bf Acknowledgements}
\end{center}

The work of E.B., M.D. and A.P. was partially supported by RFBR-DFG
grant 00-02-04011, RFBR grant 01-02-16710, scientific program 
"Universities of Russia"
grant 990588, CERN-INTAS grant 99-0377 and INTAS grant 01-0679. 
E.B. and M.D. thank very much DESY-Zeuthen for hospitality.

\end{document}